\documentclass[num-refs]{wiley-article}
\usepackage{graphicx} 
\usepackage{comment} 
\usepackage{float} 
\usepackage{wrapfig} 
\usepackage{hyperref}
\usepackage{multicol}
\papertype{Original Article}
\title{Analysis of Code and Test-Code generated by Large Language Models}

\author[1\authfn{1}]{Robin Beer}
\author[1\authfn{1}]{Alexander Feix}
\author[1\authfn{1}]{Tim Guttzeit}
\author[1\authfn{1}]{Tamara Muras}
\author[1\authfn{1}]{Vincent Müller}
\author[1\authfn{1}]{Maurice Rauscher}
\author[1\authfn{1}]{Florian Schäffler}
\author[2\authfn{1}]{Welf Löwe, PhD}

\contrib[\authfn{1}]{Equally contributing authors.}

\affil[1]{Technical University of Applied Sciences Augsburg, Augsburg, Germany}
\affil[2]{Linnaeus University, Kalmar/Växjö, Sweden}

\corraddress{Welf Löwe, Linnaeus University, Sweden}
\corremail{Welf.Lowe@lnu.se}
\fundinginfo{Linnaeus Excellence Center on Data Intensive Sciences and Applications, DISA}
\runningauthor{Beer, Feix, Guttzeit, Muras, Müller, Rauscher, Schäffler, Löwe}

\begin{document}

\begin{frontmatter}
\maketitle

\begin{abstract}
Large language models (LLMs), such as ChatGPT and Copilot, are transforming software development by automating code generation and, arguably, enable rapid prototyping, support education, and boost productivity. Therefore, correctness and quality of the generated code should be on par with manually written code. To assess the current state of LLMs in generating correct code of high quality, we conducted controlled experiments with ChatGPT and Copilot: we let the LLMs generate simple algorithms in Java and Python along with the corresponding unit tests and assessed the correctness and the quality (coverage) of the generated (test) codes. We observed significant differences between the LLMs, between the languages, between algorithm and test codes, and over time. The present paper reports these results together with the experimental methods allowing repeated and comparable assessments for more algorithms, languages, and LLMs over time.
\end{abstract}
\end{frontmatter}

\section{Introduction}
Artificial Intelligence (AI) has already established a significant role in modern software development. AI-powered coding assistants of various kinds have been introduced including tools for bug and vulnerability detection, program analysis, test automation and fuzzing, and code generation. Due to the popularity and improvements of Large Language Models (LLMs), fully automated code generation seems to be in reach. Expectations range from an increase in programming productivity, over lifting the abstraction level in programming, to the end of programming as a profession, substitute by prompts that describe what software is supposed to do instead of how.

This raises several questions: How advanced are LLMs today, i.e., can they generate correct and clean code and can they generate test code to provide confidence in the correctness? Also, are their capabilities different for different programming languages and, finally, are they improving over time? Our study aimed at investigating these questions in controlled experiments.  

However, our study is restricted in several ways. First, we selected basic textbook algorithms, published in many textbooks and repositories. This setup should be in favor of LLMs as they learn from examples. Second, we restricted ourselves to Python and Java ranking second and third in the top programming languages on GitHub~\cite{TopLang2023}, which again was in favor of LLMs. The reason for not choosing JavaScript topping the ranks is that we wanted to include a compiled, typed language (Java) together with an interpreted, untyped language (Python). Third, we analyzed the code generation performance of only two LLM tools, ChatGPT \cite{OpenAI2023} and GitHub Copilot \cite{Github2023}, representing general purpose (ChatGPT) and programming specific (Copilot) LLMs; we acknowledge that other LLM code generation tools might perform differently, cf. a selection of such tools~\cite{CGtools2023}.

\subsection{Background}
The code generating AI systems considered in the present study are based on Large Language Models. 

A \textbf{Large Language Model} is a type of artificial intelligence used for processing language \cite{du_shortcut_2023}. LLMs are machine learning models, more specifically deep neural networks, and a subset of language models, or language modelling, which aims to improve the ability of machines to understand language. Typically, this is done by modelling the probabilities of word sequences in text, and thus calculating the likelihood of these sequences occurring in the rest of the text \cite{chang_survey_2023, douglas_large_2023}. The distinction between conventional language models and LLMs is the amount of mostly unlabeled data used for training. The capacity of LLMs is scaled with the size of the training data set \cite{douglas_large_2023}.

\textbf{ChatGPT} \cite{OpenAI2023} is an LLM based chat system created by OpenAI; it is a fine-tuned variant of the LLM GPT-3.5. A user can access the system by texting with a chatbot via a web interface or an API. The output has been optimized to simulate human interaction with the implementation of human feedback and correction \cite{chatgpt_explanation}. It is not specifically trained for code generation. GPT is an abbreviation for "Generative Pre-trained Transformer", which nowadays represents a set of LLMs developed by OpenAI. \textbf{Transformer} (deep machine learning) models rely on a mechanism mimicking cognitive attention~\cite{Vaswanietal2017}. The quantity of training data and the number of parameters for the models have increased with each version \cite{lm_unsupervised, lm_few_shot}. However, official information on GPT-4's parameter count is not yet available. GPT-3 was never officially released, but was instead offered via an API to restrict usage of the model\cite{gpt-3}. ChatGPT can currently be used with both GPT-3.5 and GPT-4. However, access to the latter requires payment \cite{chatgpt-4_payment}.

\textbf{GitHub Copilot} \cite{Github2023} is an LLM based coding assistant designed to enhance developers' productivity by enabling them to write code fast and with little effort. It retrieves data from comments and code to propose complete functions or individual lines of code. GitHub Copilot is based on a generative AI model developed by GitHub, OpenAI and Microsoft, trained on publicly available natural language and source code samples, including public repository code on GitHub. Unlike ChatGPT, Copilot is a built-in extension to the IDE or editor \cite{copilot_explanation}. 

AI systems that generate code have been researched earlier and mainly
addressed easy tasks, such as generating one line functions or simple algorithms \cite{Chenetal2021,HE2023}.

\subsection{Problem Formulation}
This study addresses three main questions.
\begin{enumerate}
    \item What is the current capacity of LLM based tools to generate correct and clean code?
    \item Is there a significant difference between different LLM tools? As we extend the work documented in \cite{HE2023}, this also includes the question of their improvements over the last year.
    \item Is there a significant difference between the programming languages, the code is generated for?
\end{enumerate}

Given the huge interest in and the fast advancement of the assessed tools, the answers to the above questions can only be a snapshot and are, hence, expected to expire rather sooner than later. Therefore, the foremost problem to be solved by our work is to provide an evaluation framework---including benchmark problems, metrics, and statistical tools---to reevaluate the questions in a longitudinal study to see trends and to speculate about future opportunities of LLM based code generation in a more profound way. 

\subsection{Structure}
The remainder of the paper is structured as follows: Section~\ref{sec:method} introduces the scientific method and experimental setup for addressing the problem and answering the above questions. Sections~\ref{sec:experiment} and ~\ref{sec:eval} describe details of the conduction and the results of the experiment, resp. Section~\ref{sec:discussion} discusses the results, answers the research questions, and lists possible threads to validity of these answers. Section~\ref{sec:related} compares related studies with ours. Finally, Section~\ref{sec:conclude} concludes the paper and shows possible directions of future work.

\section{Method}\label{sec:method}
To assess the Large Language Models' code generation performance, we instructed them to generate code in multiple programming languages. Such code was either an algorithm implementation or a unit test suite for the same. The resulting generations were then evaluated using various metrics to assess the code correctness and quality answering the following concrete questions:
\begin{enumerate}
\item How well does ChatGPT (GitHub Copilot) provide \textbf{correct code} in Python (Java) from specified instructions?
\item How good is ChatGPT (GitHub Copilot) at generating Python (Java) \textbf{code of high quality}?
\item How well does ChatGPT (GitHub Copilot) provide \textbf{correct Unit Tests} in Python (Java) from specified instructions?
\item How good is ChatGPT (GitHub Copilot) at generating Python (Java) \textbf{Unit Tests of high quality}?
\item Can we observe significant \textbf{differences} between \textbf{Java and Python}?
\item Can we observe significant \textbf{differences} between \textbf{ChatGPT and GitHub Copilot}?
\item Can we observe significant changes compared to the baseline study~\cite{HE2023}, \textbf{differences over time}?
\end{enumerate}

Implementing the infrastructure for answering the questions, provided the evaluation framework that we suggest for future repetitions of the assessment in a longitudinal study.

\subsection{Algorithm Selection}
We have chosen twelve algorithms that the LLMs are supposed to generate. Since the study of Hansson and  Ellréus~\cite{HE2023} served as the baseline for our experiments, we selected the same six algorithms. They started with the first five algorithms described in \cite{Cutajar2018}. In order to not argue about any bias in this selection, we included the missing algorithms of the top ten listed in this article. Following the arguments in \cite{HE2023} for including Binary-to-Decimal conversion to increase the diversity in the set of algorithms, we added the Egyptian Fractions algorithm. This way we have at least two different examples for any type of algorithm in our set that includes the following twelve algorithms (type of algorithm in parentheses):
\begin{itemize}
    \item Bellman-Ford$^*$, (graph, optimization algorithm)
    \item Binary Search$^*$, (search algorithm)
    \item Binary To Decimal$^*$, (number encoding algorithm) 
    \item Breadth First Search (BFS)$^*$, (search algorithm)
    \item Depth First Search (DFS)$^{**}$, (search algorithm)
    \item Dijkstra$^{**}$, (graph, optimization algorithm)
    \item Egyptian Fractions$^{***}$, (number encoding algorithm)
    \item Floyd-Warshall$^{**}$, (graph, optimization algorithm)
    \item Knapsack$^*$, (optimization algorithm)
    \item Kruskal$^{**}$, (graph, optimization algorithm)
    \item Merge Sort$^*$, (sorting algorithm)
    \item Quick Sort$^*$, (sorting algorithm)
\end{itemize}
In short, they were selected because they $^*$ were contained in \cite{HE2023}, $^{**}$ fill the gaps of the top ten algorithms in~\cite{Srivastava2023}, and $^{***}$ were needed to have at least two different number encoding algorithms.

\subsection{Formulating Prompts}\label{sec:prompts}

Both LLMs are instructed with prompts. Depending on the amount and type of information in a prompt, the models generate different results. We created individual prompts for each combination of programming language and algorithm. 

\subsubsection{Source Code Generation Prompts}
Generally, all source code prompts contain the algorithm, the class name, and detailed information describing the class structure. This includes inner classes, constructors and method declarations with their access modifiers, return values and parameters. The syntax for the class structure given for Java was sufficient for the AI to identify the intended programming language. This did not continue to hold for Python. Therefore, we included the programming language in the prompt. Below, an example of two prompts for Java and Python, respectively, to generate a binary search algorithm implementation. 
\begin{verbatim}
// Implement a non-static class named BinarySearch.
// Implement the public binarySearch(int, int[]) method. The method should return a boolean.

# Implement a non-static class named BinarySearch in python.
# Implement the public binary_search() method.
# The method should use an integer, an integer list as parameters.
# The method should return a boolean.
\end{verbatim}
All prompts used can be found in the appendix. 

\subsubsection{Test Case Generation Prompts} \label{sec:testgenapp}
For the unit test case generations, three different approaches were used to prompt the LLMs; all yielded promising results in preliminary tests. Documentation of these tests can be found in the source code repository in Appendix~\ref{app:repos}.

The \textbf{first approach} provides only the structure of the algorithm in text form, including the names of the variables and the functions, similar to the prompt generating the algorithm source code. The Binary Search unit tests prompt in Python is provided as an example.
\begin{verbatim}
# Implement a python test class called BinarySearchTest using the unittest module.
# The class BinarySearch that is to be tested contains the following methods:
# A public method binary_search(), which takes an integer and an integer list as parameters and 
  returns a boolean.
# Implement at least 3 different test cases.
\end{verbatim}

The \textbf{second} and  \textbf{third approach}  provides a source code implementation of the algorithm to be tested, complemented with a short comment requesting unit tests for the code. For instance, an example comment for the Binary Search unit test in Python is shown below:
\begin{verbatim}
# Implement a test class for the binary search algorithm above using the python unittest module.
# Implement at least 3 different test cases.
\end{verbatim}
The difference between the second and the third approach was the actual source code provided. In the \textbf{second  approach}, the manually written algorithm reference implementation was provided. It was the same implementation as used for validating the functionality of the respective unit tests for that algorithm. 

The supplied code in the \textbf{third approach} was a randomly selected generated implementation for the respective algorithm, taken from the first part of our experiment. The selected implementation, however, needed to pass the manually written reference unit tests as a prerequisite, i.e., the unit tests used for validating the source code generation for that algorithm.

All prompts used in the experiments are listed in the Appendix~\ref{app:prompts}. 

\subsection{Generating Code}

We generated code with \textbf{ChatGPT} by using its web interface. ChatGPT is set up as a chatbot and offers multiple chat windows with different conversations. Since this LLM uses the existing conversation in the current window as the context for its next generation, a new window was required for each generation. This step ensures that the individual code samples remain independent and unbiased. A second way of preventing different code generations from affecting each other is to disable the “Chat history \& training” feature in the Data Controls setting. This will result in a single chat window that can be cleared, ensuring that prior content will have no effect on new generations. 

The previously formulated prompt, see Section~\ref{sec:prompts}, was entered into the chat input box. ChatGPT  then generated a detailed answer containing an implementation of the desired algorithm. However, it tended to generate a main function or even an entire main class. Due to the significance of the lines produced in the later analysis, we decided to remove the extra code to enable comprehensive evaluation of ChatGPT's ability to generate the desired algorithms with high quality. During the process of generating test cases, ChatGPT sometimes generated an implementation of the algorithm itself. Those implementations were also removed.

As a built-in tool for IDEs and editors, \textbf{GitHub Copilot} cannot generate code in the same general way as ChatGPT and required some extra steps. It uses the contents of the current working directory as context for generation. Therefore, the independence of each sample required the creation of a fresh Java project and an empty folder containing a Python file, resp. The code generation files had to be named after the chosen algorithm, written in the appropriate casing convention. Prior to conducting the generation process, the file had to include the prompt as a comment, followed by the class body with the cursor positioned within it. In certain IDEs and editors, GitHub Copilot provides the option to generate up to ten samples simultaneously in a separate tab or window. As this feature significantly streamlined the generation process; we decided to take advantage of it. Since accepting one of the options would have been taken into account for future generations in this project, we only copied each generation into a new file in the experiment workspace.

\subsubsection{Manual Correction of Source Code Generations} \label{sec:manCorrectionSource}

GitHub Copilot is, by design, unable to generate lines of code outside the current scope. This led to some missing import statements. Without adding these manually, the results of the experiment would have been significantly affected. Therefore, we decided to include them using the refactoring functionality of the IDEs. This made the results of ChatGPT and GitHub Copilot more comparable.

In the event that the code was not able to be compiled or interpreted, we opted to regenerate the sample using the same method as previously outlined in order to achieve an adequate sample size.

\subsubsection{Manual Correction of Test Case Generations} \label{sec:manCorrection}

For both Java and Python test case generation, our analysis required the code to be compilable or, in the case of Python, interpretable. Consequently, in instances where this was not the case, the generated code had to be manually corrected. The sole intention of these corrections was to make the code syntactically correct with as few modifications as possible. To ensure this, we decided to do the corrections in pair programming. 

The only exception in this case was the treatment of uncompleted assertion statements. Since filling these statements with syntactically but not semantically correct content would have affected the results of later checks of the code, we decided to remove uncompleted assertion statements.

\subsection{Evaluation Metrics} 
We defined different code correctness and quality metrics in order to evaluate the generated code, which will be described below.

\subsubsection{Code Correctness}\label{sec:code_correctness} 
To verify that the code generated by the LLMs is correct, we integrated several reference unit tests into our project. These implementations reuse by-and-large unit tests from third-party sources. To ensure that they are implemented correctly, we tested them against reference implementations of the algorithms, reused from third-party sources as well.\footnote{We named the third-party (test) sources in our repository for the respective algorithm implementation and we documented the changes made to them. See for:\\
Java: \url{https://github.com/tguttzeit/AI-Code-Examination/tree/main/java_app/src/main/java/BookExamples},\\
Java tests \url{https://github.com/tguttzeit/AI-Code-Examination/tree/main/java_app/src/test/java/BookExamples},\\
Python \url{https://github.com/tguttzeit/AI-Code-Examination/tree/main/py_app/src/BookExample},\\
Python tests \url{https://github.com/tguttzeit/AI-Code-Examination/tree/main/py_app/test/BookExample}.} All reference unit tests defined for an algorithm must pass for a generated code to be considered correct.

To capture the correctness of generated unit tests, we tested their code against the above-mentioned reference implementations of each algorithm. We manually checked that the reference implementation was correct and passed all reference unit tests. For the generated test code to be considered correct, all its unit tests must pass the reference source code implementation.

\subsubsection{Code Quality} \label{sec:code-quality}
To measure the quality of the generated code in addition to its correctness, we've established guidelines to follow for high-quality code. For the programming language Java, we've adopted quality rules from the referenced bachelor thesis by Hansson and Ellréus~\cite{HE2023}, which have originally been derived from the book \textit{Clean Code} by Robert C. Martin~\cite{Martin2008}. The principles used for Java are following:
\begin{itemize}
    \item The files should not be over 500 lines long.
    \item A line should not be more than 120 characters long.
    \item Magic numbers should be hidden behind constants.
    \item Functions should not be more than 20 lines long.
    \item Functions should not have more than three arguments.
    \item There should not be nested loops of a depth of more than one level.
    \item There should not be more than one statement per line.
    \item There should not be any inner assignments~\cite{HE2023}
\end{itemize}

For Python, we followed the guidelines outlined in the PEP-8 Style guide to ensure referring to appropriate standards. 
We therefore implemented the following quality rules:
\begin{itemize}
    \item A line should not exceed 79 characters.
    \item A file should not be longer than 500 lines.
    \item Magic values should be hidden behind constants.
    \item Functions and methods should have a maximum of three arguments.
    \item Nested blocks should not contain more than two blocks.
    \item There should be only one statement per line.
    \item Function and method names should not exceed 20 characters.~\cite{PEP2001}
\end{itemize}

According to these guidelines, we counted both the number of rule violations and the number of lines generated for each code sample. Code Quality is determined by the ratio of violations to the number of generated lines. 

We only evaluated the code quality of the generated source code, not of the generated unit tests. Because we had to modify some of the unit test code as described in Section~\ref{sec:manCorrection}, which would have influenced the results, we decided that code coverage of the generated tests was a more instructive quality  metric on test code.

\subsubsection{Code Coverage} 

As code quality metric for generating tests, we measured the test coverage of the source code. We analyzed  the coverage of instructions and branches in combination, as described in more detail in Section~\ref{sec:testCaseGen}.

\subsubsection{Modification Rate of Generated Test Code} 

Besides the correctness and coverage of the generated tests, we evaluated the number of required modifications; recall the modifications explained in Section~\ref{sec:manCorrection}. These were quantified using the Levenshtein distance, a metric for measuring the difference between two strings, here the generated and the manually modified test code (strings). It is informally defined as the minimum number of edit operations (insertions, deletions or substitutions of single characters) required to change one string into the other~\cite{Levenshtein1966}. To calculate the modification rate, we divided the Levenshtein distance by the number of characters in each code sample, excluding the length of imports and comments.

\subsection{Automated Code Evaluation} \label{sec:automatedPipelines}

\subsubsection{Source Code Evaluation}
The generated code was evaluated for the metrics explained above. The evaluation scripts were written in JavaScript, while the automation of those scripts was implemented as a pipeline in Bash.

The assessment of code quality relied primarily on the use of the linting tools \textit{Checkstyle} for Java and \textit{Pylint} for Python. We configured both tools to detect only those quality defects that violate our defined quality rules. A separate script counts the lines of code in each file, ignoring any type of comment and imported package, as the number of lines for each file is also needed in the later analysis.
We opted to omit import statements, since importing complete libraries and packages can be substituted by importing only necessary implementations. This may result in numerous import lines instead of one, which could have influenced the outcomes.

Code correctness was checked using reference unit tests from third-party sources; we refer to the discussion in Section \ref{sec:code_correctness} and the documentation in the \verb|BookExample| directories on GitHub \url{https://github.com/tguttzeit/AI-Code-Examination}. To guarantee the consistency of the test suite with the generated code, we incorporated the necessary code structure into the prompt, as previously explained. We utilized the JUnit framework for Java and the unittest.py package for Python code, resp.

\subsubsection{Test Case Evaluation} \label{sec:testCaseGen}
Similar to the experiment for the generated source code, the evaluation of the generated test codes was automated using scripts as much as possible. The pipeline for the test case generation experiment had to be split into two stages, allowing for the explained manual changes in between.

In the first stage after generating the code, a script had to be run to detect any file that could not be compiled or interpreted. If checking Java files, the script attempts to compile each generated test and copies any file that could not be compiled into a designated directory. As Python code does not require compilation, we took advantage of Pylint's ability to detect any error in the source code prior to execution. Any file discovered to contain errors was copied to a separate directory, similarly to the approach in Java. Additionally, both scripts produce a JSON file that enumerates all files requiring correction. These corrections were carried out on the initial document, and the measure of the changes between the two files was then calculated using Levenshtein distance.

The second stage of the pipeline starts by running the generated unit tests to verify their correctness. As explained above, we exclusively utilized source code confirmed functional. Therefore, a correctly generated test should not result in failure. Just as in the source code generation experiment, we used the JUnit framework for Java and the unittest.py package for Python unit tests.

We incorporated a code coverage analysis to our pipeline as a final check on the quality of the generated tests. For the Python code, we used the coverage.py package; for the Java code, we used  the JaCoCo library for coverage analysis. In contrast to coverage.py, JaCoCo's default report provides coverage details for the test class, the source code class, and its inner classes. For all examined coverage metrics, we calculated the mean of the source code class and its inner classes, excluding the test class.

\subsection{Statistical Analysis}

We used Matlab and Python libraries and the AI-Therapy Statistics tool to analyze the data provided by the pipelines. To visualize the results in the form of graphs, we developed MATLAB scripts. As the data is separated per algorithm in the respective JSON files, we first merged the data with the scripts before calculating the frequencies and creating all graphs provided in Section~\ref{sec:experiment}. 

For descriptive statistics and to detect significant differences within the data, we used the web-based tool AI-Therapy Statistics that provides statistical functions for data analysis. Appendix~\ref{app:statistical-analysis} contains the links to all analyses performed, including mean, mode, median, and dispersion for all the metrics mentioned above. A normality test was performed on each dataset before proceeding with hypothesis testing. As the datasets did not follow a normal distribution, we used non-parametric tests. To compare two groups, e.g., for the programming languages Java and Python or for the AI models ChatGPT and GitHub Copilot, we used the 
Mann-Whitney U test\footnote{\url{https://en.wikipedia.org/wiki/Mann-Whitney_U_test}}. This test uses mean ranks or medians to compare the scores of two groups on a variable. To test for significant differences between the three approaches to generating test cases, we used the Kruskal-Wallis test\footnote{\url{https://en.wikipedia.org/wiki/Kruskal-Wallis_one-way_analysis_of_variance}} that can compare more than two groups on a variable. As the Kruskal-Wallis test is an extension of the Mann-Whitney U test, it is also based on mean ranks or medians~\cite{VA2019}.

For analyzing the differences in correctness and quality over time, i.e., between Spring 2023 ($t_0$) and Fall 2023 ($t_1$), we first apply the Mann-Whitney U test again. However, the samples generated at $t_0$ and $t_1$ are arguably not independent. Therefore, we analyze pairs of outcome at $t_0$ and $t_1$ for the same algorithm when measuring correctness (quality) of ChatGPT (Copilot) generated codes, respectively. We calculate the means of the outcomes for each algorithm over its $50$ generations separately at $t_0$ and $t_1$ and assess whether these population means differ using paired difference tests. 
As the differences may not be normally distributed, we apply the (non-parametric) Sign test\footnote{\url{https://en.wikipedia.org/wiki/Sign_test}} and, as this test may find the differences falsely non-significant, we crosscheck with the (also non-parametric) Wilcoxon signed-rank test\footnote{\url{https://en.wikipedia.org/wiki/Wilcoxon_signed-rank_test}}. 

\section{Conducting the Experiment}\label{sec:experiment}

According to the methodology described previously, we conducted the experiment.
Using the prompts and instructions described above, we generated code samples with ChatGPT and GitHub Copilot. We generated source code and unit tests separately for both Java and Python. An equal number of samples was produced for both programming languages. The precise distribution of the sample size is outlined below.

For the purpose of evaluating source code, we generated 50 code samples for each algorithm using ChatGPT and GitHub Copilot. As we have selected twelve algorithms, this results in a combined sample size of 600 per AI model and 1200 code samples overall for each programming language.

To evaluate the test code, we generated 10 samples for each of the three approaches, resulting in 30 unit test samples for each combination of algorithm and AI model. As a consequence, there are 360 samples per AI model for twelve algorithms and 720 code samples in total for each programming language. 

After generating, we added all code samples to our project and executed the two pipelines described in Section~\ref{sec:automatedPipelines}. To analyze the provided data, we transferred the resulting JSON files to our analysis repository. The files can be accessed and downloaded from our repository linked in appendix~\ref{app:repos} to reproduce our findings. For the analysis, we used the tools mentioned above, generated graphs, and evaluated differences between the AI models and programming languages, which will be described in the following.

\section{Evaluation of Examination Results}\label{sec:eval}

The experiment was divided into source code generation and test case generation, as previously mentioned. Source code generation analyzed the generated implementations of an algorithm according to instructions provided in the prompt, while test case generation generated unit tests for the algorithm.

In the following sections, we explain the results of our research. The results for all the metrics are presented with graphs that illustrate the relative outcomes for each metric. Tables display the pure outcomes for the metrics (absolute results) and can be found in the appendix, together with the statistical analysis results including the detailed results of the hypothesis tests.

Further results and diagrams including deeper analysis on the topics stated below can be found in our repository on GitHub.

\subsection{Source Code Generation}

Firstly, we examined the code correctness and quality of the generated source code for both programming languages and all algorithms.

\subsubsection{Code Correctness}

\begin{figure}[ht]
    \centering
    \begin{minipage}[c]{.5\textwidth}
        \centering
        \includegraphics[width=1\linewidth]{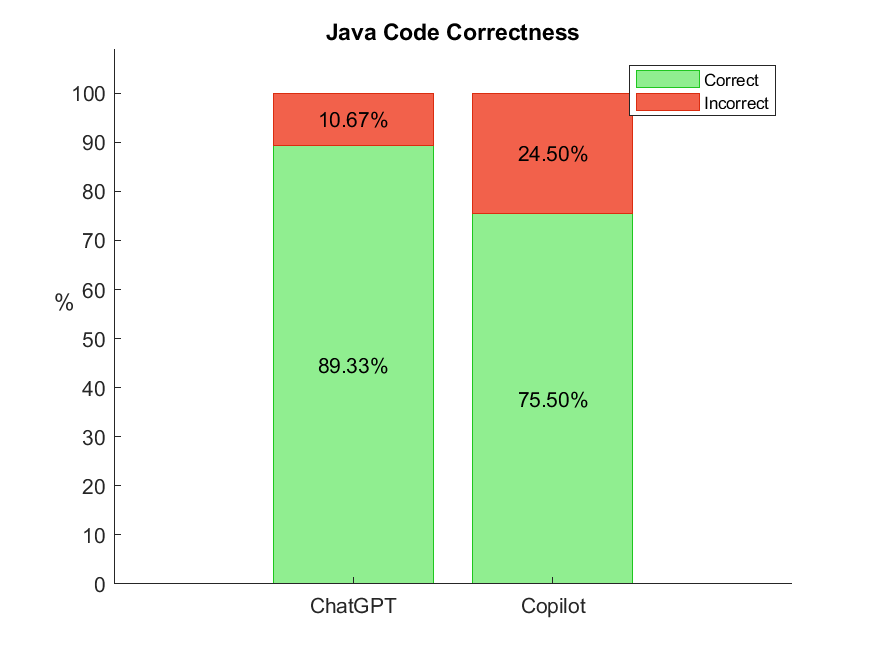}
        \captionsetup{justification=centering}
        \caption{Code Correctness results for Java}
        \label{fig:scg-correctness-java}
    \end{minipage}
    \begin{minipage}[c]{.49\textwidth}
        \centering
        \includegraphics[width=1\linewidth]{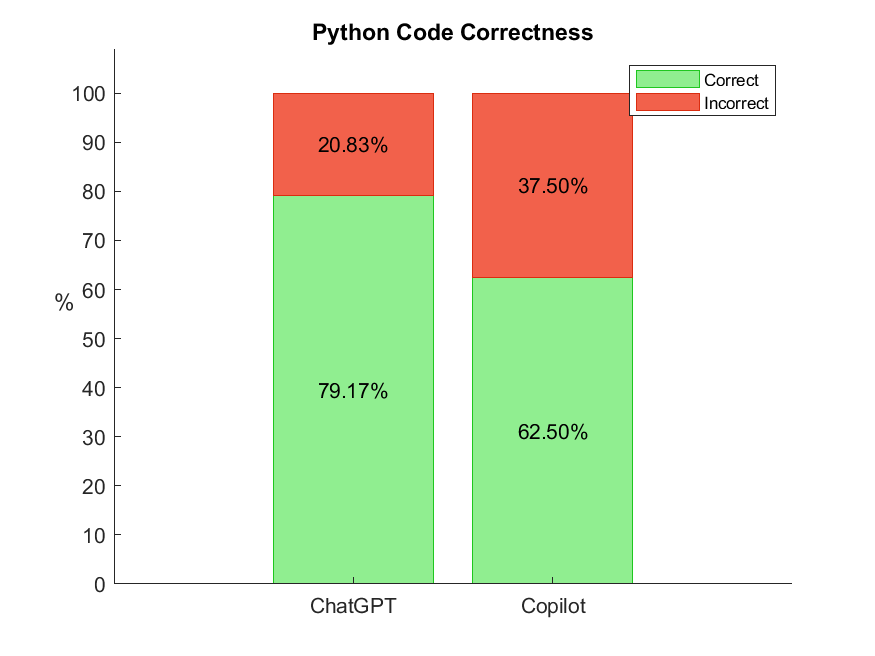}
        \captionsetup{justification=centering}
        \caption{Code Correctness results for Python}
        \label{fig:scg-correctness-python}
    \end{minipage}
\end{figure}

\paragraph{Java}
Generating source code for twelve algorithms in Java yields the following outcomes for the code correctness: Both AI models can generate correct Java code. There is a noticeable difference between the models, with ChatGPT outperforming GitHub Copilot by about 14~\%, see Figure~\ref{fig:scg-correctness-java}. This can be further demonstrated through a hypothesis test. The Mann-Whitney U test reveals a statistically significant difference between the two at a significance level of 0.01.

\paragraph{Python}
The results for generating source code in Python are as follows:
ChatGPT and GitHub Copilot are capable of producing valid Python code. There is a clear difference between the AI models again, with ChatGPT performing better with approximately 17~\%, see Figure~\ref{fig:scg-correctness-python}. Comparing the two models using the Mann-Whitney U test also indicates a statistically significant difference at a significance level of 0.01.

\paragraph{Comparing Java and Python}
As previously seen, the results for code correctness in Java and Python are similar. The AI models produce accurate code in both languages, although the results for Java show a clear advantage over Python. For ChatGPT, there is roughly a 10~\% difference comparing (the correctness of) Java and Python codes and for GitHub Copilot, a 13~\% difference comparing Java and Python. 

When analyzing the ChatGPT correctness results of the two programming languages, the Mann-Whitney U test reveals a statistically significant difference between Java and Python, based on a significance level of 0.01. Doing this for the correctness results of GitHub Copilot yields the same outcome.
However, the \emph{differences} between ChatGPT and GitHub Copilot within the languages are almost identical between the two, around a mean of about 14~\% in Java and 17~\% in Python. 

\subsubsection{Code Quality}

\begin{figure}[ht]
    \centering
    \vspace{-20pt}
    \begin{minipage}[c]{.49\textwidth}
        \centering
        \includegraphics[width=1\linewidth]{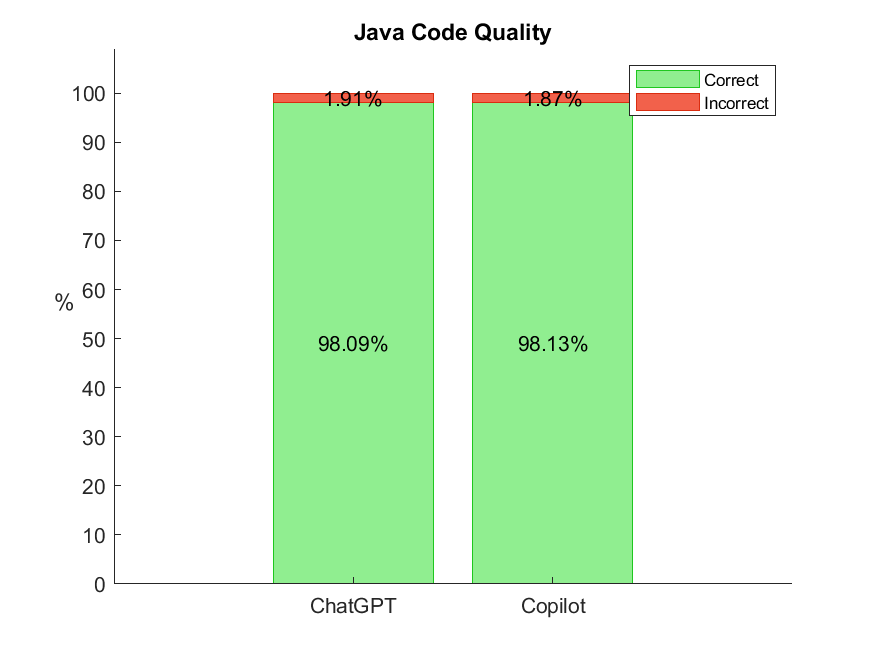}
        \captionsetup{justification=centering}
        \caption{Code Quality results for Java}
        \label{fig:scg-quality-java}
    \end{minipage}
    \begin{minipage}[c]{.49\textwidth}
        \centering
        \vspace{11pt}
        \includegraphics[width=1\linewidth]{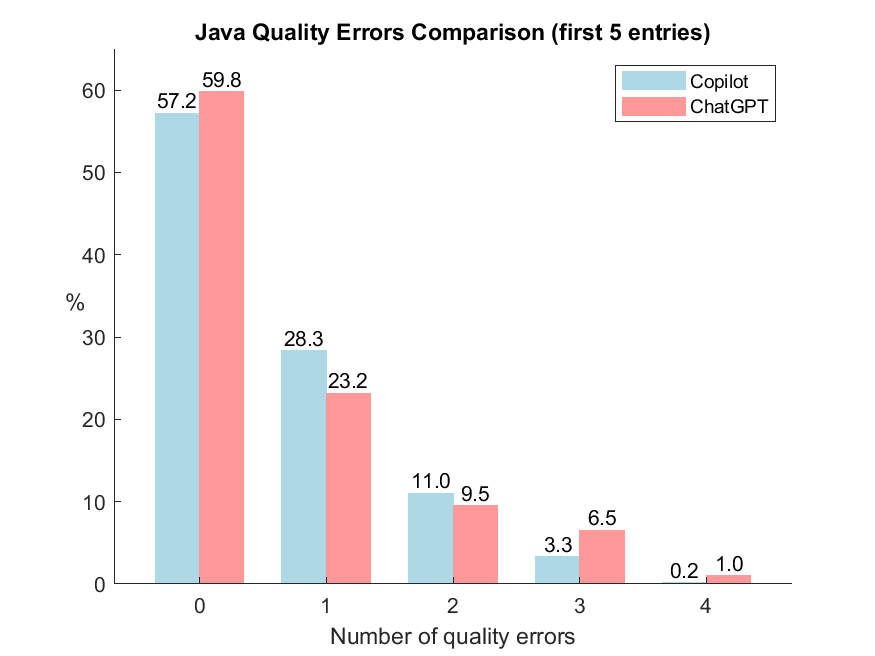}
        \captionsetup{justification=centering}
        \caption{Comparison of detected quality errors for Java}
        \label{fig:scg-quality-errors-java}
    \end{minipage}
\end{figure}

\paragraph{Java}
The quality results of the generated source code indicate that both AI models produce high-quality Java code, according to our defined quality guidelines in Section~\ref{sec:code-quality}, see Figure~\ref{fig:scg-quality-java}.
GitHub Copilot and ChatGPT do not differ in terms of code quality, as determined by a Mann-Whitney U test at a significance level of 0.05.

Comparing the number of quality violations detected for each algorithm generation, both AI models mostly produce implementations free from violations. In general, there is no large difference between ChatGPT and GitHub Copilot regarding the number of quality violations within generated code. Both achieve a maximum of 4 quality errors per implementation, see Figure~\ref{fig:scg-quality-errors-java} for the distribution of violations in the generated files.

\paragraph{Python}
\begin{figure}[ht]
    \centering
    \vspace{-20pt}
    \begin{minipage}[c]{.49\textwidth}
        \centering
        \includegraphics[width=1\linewidth]{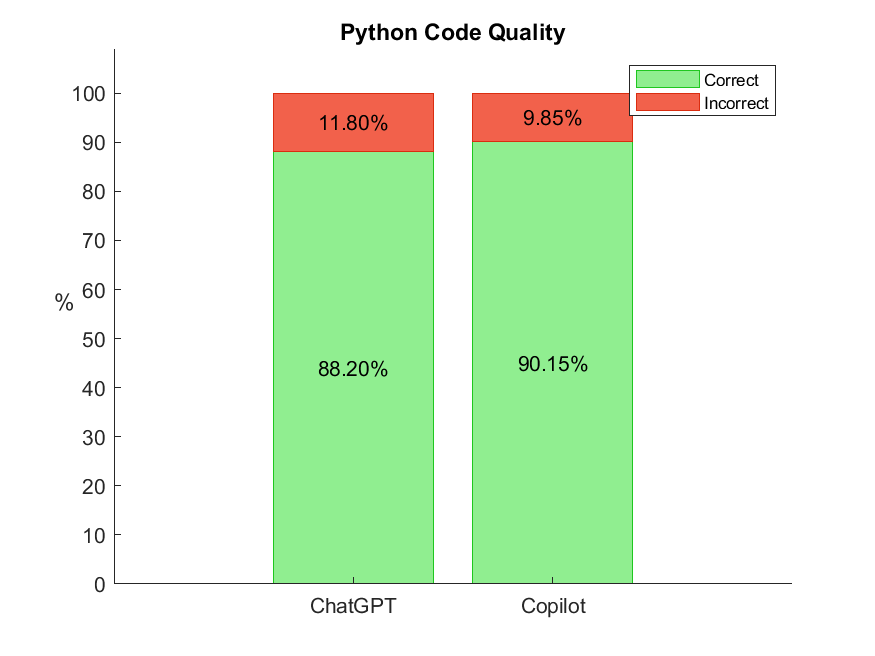}
        \captionsetup{justification=centering}
        \caption{Code Quality results for Python}
        \label{fig:scg-quality-python}
    \end{minipage}
    \begin{minipage}[c]{.49\textwidth}
        \centering
        \vspace{11pt}
        \includegraphics[width=1\linewidth]{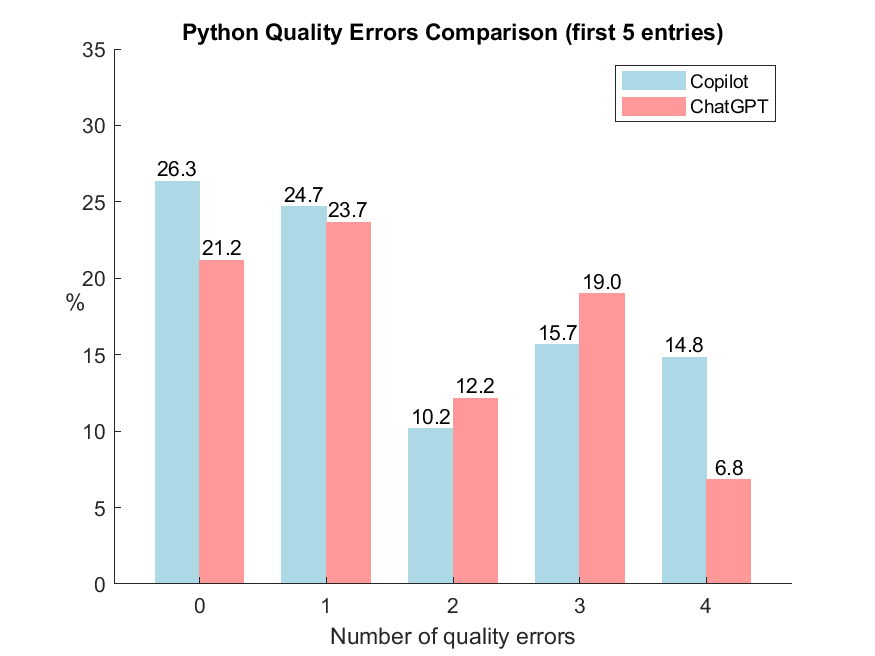}
        \captionsetup{justification=centering}
        \caption{Comparison of detected quality errors for Python}
        \label{fig:scg-quality-errors-python}
    \end{minipage}
\end{figure}
Looking at the results of generated Python code, we observe a high level of quality for both AI models. The performance of ChatGPT and Copilot is comparable, with both achieving approximately 90~\%, see Figure~\ref{fig:scg-quality-python}. There is a slight discrepancy of about 2~\%, with GitHub Copilot reaching the higher score. This difference can also be determined from the results of the Mann-Whitney U test. There is a statistically significant difference in code quality between the AI models at a significance level of 0.05. 

Regarding the number of quality violations, we observe the distribution shown in Figure~\ref{fig:scg-quality-errors-python}. It is difficult to identify a parametric distribution for the quality errors produced by GitHub Copilot. Both models generate about 50~\% of their Python codes free from errors or only containing one, but there are also a noticeable number of files with several violations. For ChatGPT, the maximum number of quality errors generated per implementation is 11, while it is 9 for Copilot.

\paragraph{Comparing Java and Python}
ChatGPT and GitHub Copilot generate high-quality code in both programming languages. As already seen in code correctness, Java code has higher code quality overall. The scores for Java are about 10~\% higher in ChatGPT and about 8~\% in Copilot. Testing the differences between the languages separately for each model, establishes a statistically significant difference for both of them at a significance level of 0.01. Additionally, the generated implementations in Python result in more quality errors per generation compared to the Java code. 

\subsection{Test Case Generation}
After generating implementations for different algorithms, we examined the generation of test cases for given instructions or given implementations of the algorithms. 

The following sections compare three approaches for generating test cases: The first strategy, called \textit{PromptOnly}, generates unit tests without any implementation of the algorithm given in the prompt. The second approach, referred to as \textit{BookExampleCode}, generates test cases for a given correct reference implementation of the algorithm.
The last approach, named \textit{AIGenerated}, is very close to the second approach, but with the modification that it uses a verified implementation of the algorithm generated by the AI model itself. The three approaches have been detailed in Section~\ref{sec:testgenapp}.

\subsubsection{Test Code Correctness}

\begin{figure}[ht]
    \centering
    \begin{minipage}[c]{.49\textwidth}
        \centering
        \includegraphics[width=1\linewidth]{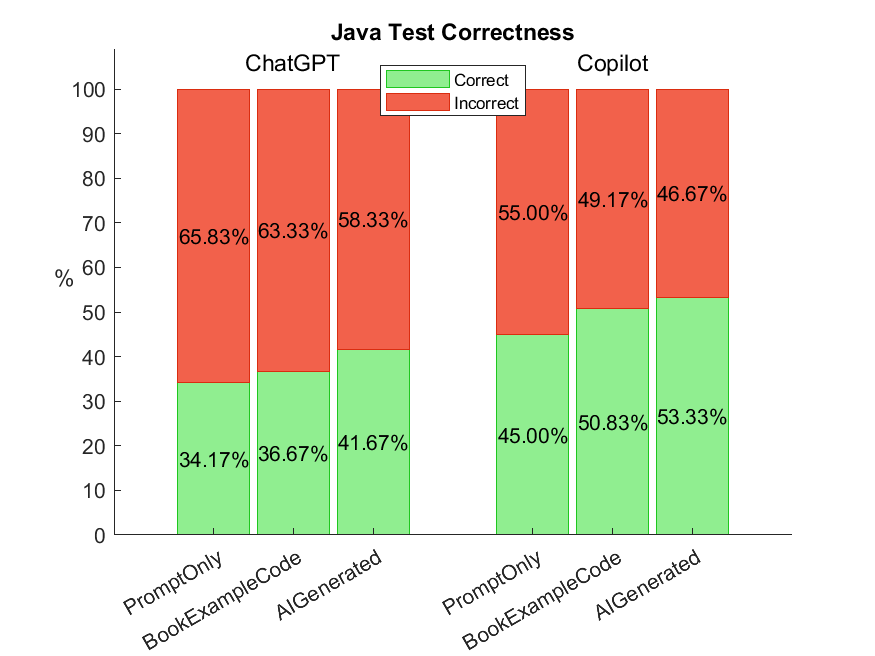}
        \captionsetup{justification=centering}
        \caption{Test Code Correctness results for Java}
        \label{fig:scg-test-correctness-java}
    \end{minipage}
    \begin{minipage}[c]{.49\textwidth}
        \centering
        \includegraphics[width=1\linewidth]{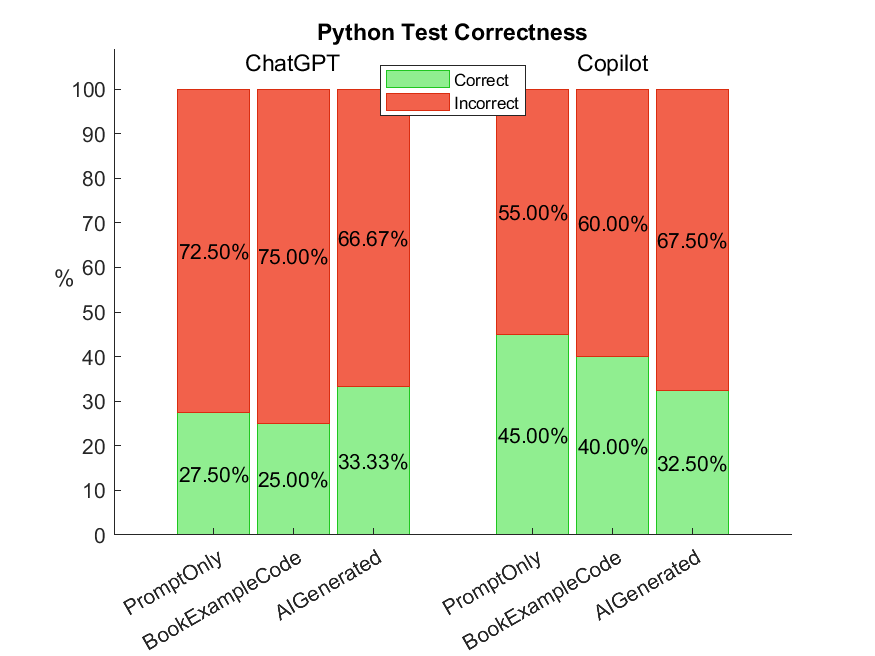}
        \captionsetup{justification=centering}
        \caption{Test Code Correctness results for Python}
        \label{fig:scg-test-correctness-python}
    \end{minipage}
\end{figure}

\paragraph{Java}
The outcomes regarding the correctness of generated Java test code are as follows:
ChatGPT and GitHub Copilot produce test code at an intermediate level, but hardly do not exceed 50~\% correctness in any approach, with Copilot performing better than ChatGPT, see Figures~\ref{fig:scg-test-correctness-java}. The Mann-Whitney U test indicates that there is a statistically significant difference between the two AI models at a 0.01 significance level.

Test case generation appears to work better with an implementation provided in the prompt, as shown by the fact that the \textit{PromptOnly} approach produces the lowest results. However, according to the Kruskal-Wallis test executed for ChatGPT and Copilot individually, the data demonstrates no statistically remarkable differences at a significance level of 0.05.

\paragraph{Python}
For Python, Copilot appears to perform better than ChatGPT, but not well either, reaching 45~\% correctness at its peak, see Figures~\ref{fig:scg-test-correctness-python}. ChatGPT on the other hand, achieves a correctness of only about 30~\% across the three approaches. The statistical analysis indicates that the difference between the two models is significant according to the Mann-Whitney U test at a significance level of 0.01. 

The performance across the approaches differs. While ChatGPT attains its highest correctness score using the \textit{AIGenerated} approach, it is the lowest for Copilot. GitHub Copilot performs best with the \textit{PromptOnly} approach. 
Regarding the performance for each AI Model, we cannot conclude any statistically significant differences between the approaches according to Kruskal-Wallis tests performed for ChatGPT and GitHub Copilot.

\paragraph{Comparing Java and Python}
Overall, the correctness results for the generated test cases are limited in both languages. The models produce a higher number of incorrect samples than correct ones. As previously observed, correct test code generation is more effective in Java compared to Python. The difference between the programming languages is statistically significant, as confirmed by the Mann-Whitney U test at a significance level of 0.01.
Concerning the variations between the AI models across the languages, Copilot performs better for both Java and Python. In Java, the models achieve comparable outcomes for all three approaches, while in Python, the results are diverse.

\subsubsection{Coverage} 
To assess the quality of test code, we use test coverage metrics, as described in Section~\ref{sec:testCaseGen}.

\begin{figure}[ht]
    \centering
    \begin{minipage}[c]{.49\textwidth}
        \centering
        \includegraphics[width=1\linewidth]{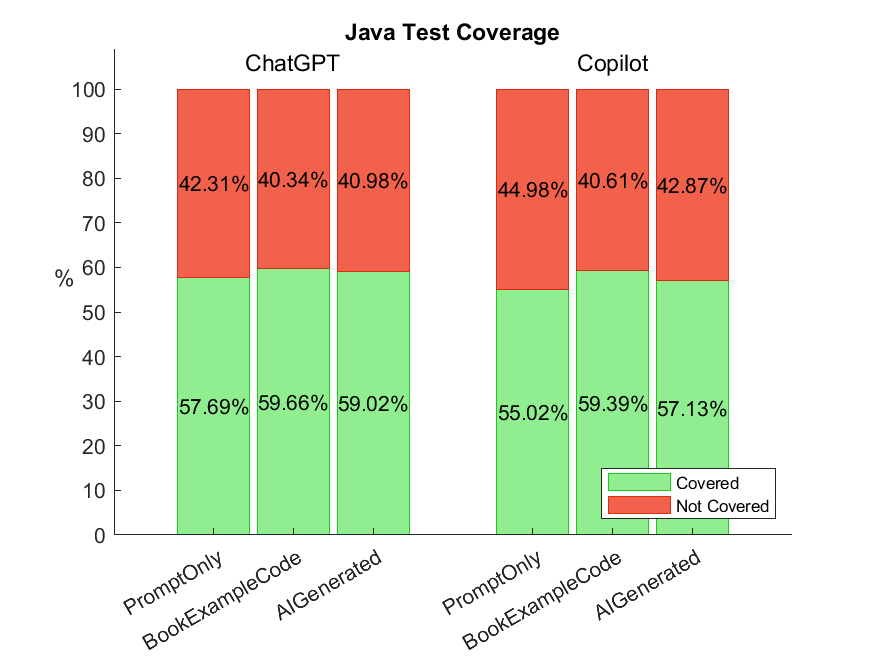}
        \captionsetup{justification=centering}
        \caption{Test Coverage results for Java}
        \label{fig:scg-coverage-java}
    \end{minipage}
    \begin{minipage}[c]{.49\textwidth}
        \centering
        \includegraphics[width=1\linewidth]{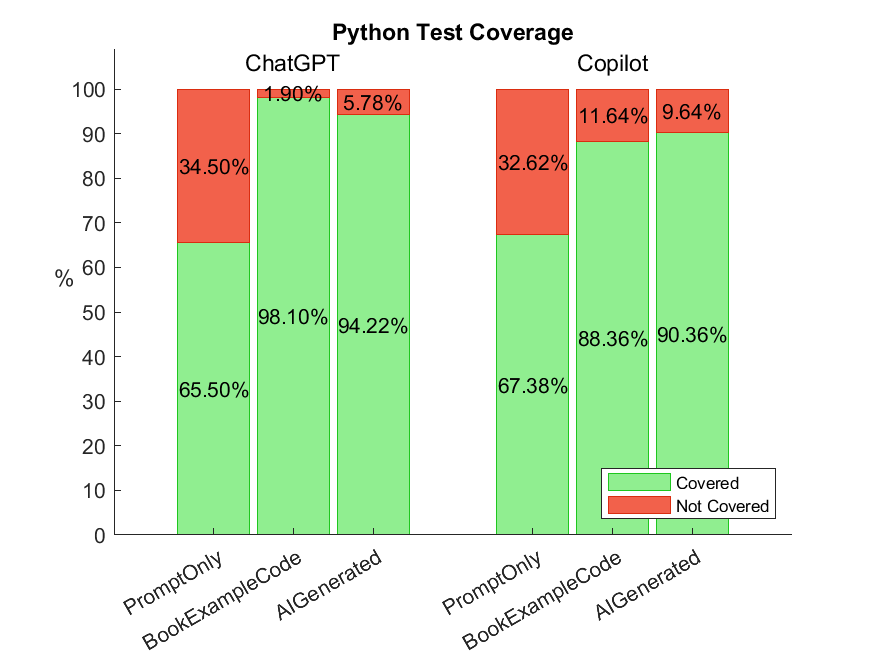}
        \captionsetup{justification=centering}
        \caption{Test Coverage results for Python}
        \label{fig:scg-coverage-python}
    \end{minipage}
\end{figure}

\paragraph{Java}
The evaluation of the coverage outcomes for generated Java unit tests indicate that across all approaches the coverage hardly reaches 60~\%, see Figure~\ref{fig:scg-coverage-java}. Although the results appear similar, using a reference algorithm implementation in the prompt (\textit{BookExampleCode}) produces the best results for generating Java test cases with ChatGPT and GitHub Copilot. There are minor differences between the approaches, with the results for Copilot showing slightly larger differences than those for ChatGPT. This is verified by a Kruskal-Wallis test performed separately for the AI models. The results reveal a statistically significant difference in the three approaches for Copilot on a significance level of 0.05, but none for ChatGPT. 

Comparing the two models shows small variations for the \textit{PromptOnly} and \textit{AIGenerated} approaches. The Mann-Whitney U test confirms that the disparity between ChatGPT and GitHub Copilot is statistically significant at a significance level of 0.01.

\paragraph{Python}
The results of the test coverage for Python unit tests generated are displayed in Figure~\ref{fig:scg-coverage-python}. Best results are achieved for the approaches where the correct source code to be tested is included in the prompt. In contrast, the \textit{PromptOnly} strategy performs much worse, with a gap of more than 20~\%. Both AI models show a noticeable difference, but it is more substantial for ChatGPT-generated code. The Kruskal-Wallis test shows that there is a statistically significant difference in the approaches for both AI models, at a significance level of 0.01. 

Overall, ChatGPT outperforms Copilot, achieving higher results. Nevertheless, according to the Mann-Whitney U test, the difference is not statistically significant.

\paragraph{Comparing Java and Python}
When examining differences between programming languages, Python code attains significantly better coverage results than Java. The maximum achieved coverage for Python is 98~\%, while Java does not even reach 60~\% coverage. The statistical significance of this discrepancy is demonstrated by the Mann-Whitney U test at a 0.01 significance level.

There is a clear difference between the prompting approaches in Python; in Java the prompting approach is less crucial. The approach \textit{BookExampleCode} yields the best results with only a slight advantage in Java, while in Python there is a noticeable gap between \textit{PromptOnly} and the other two approaches.

\subsubsection{Modification Rate}
\begin{figure}[ht]
    \centering
    \begin{minipage}[c]{.49\textwidth}
        \centering
        \includegraphics[width=1\linewidth]{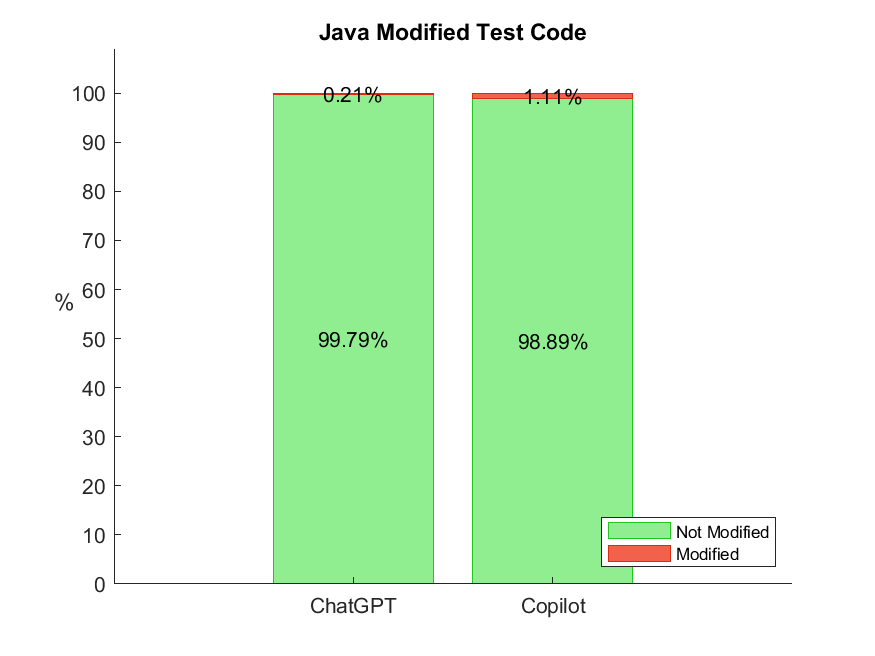}
        \captionsetup{justification=centering}
        \caption{Modification rate results for Java}
        \label{fig:scg-modrate-java}
    \end{minipage}
    \begin{minipage}[c]{.49\textwidth}
        \centering
        \includegraphics[width=1\linewidth]{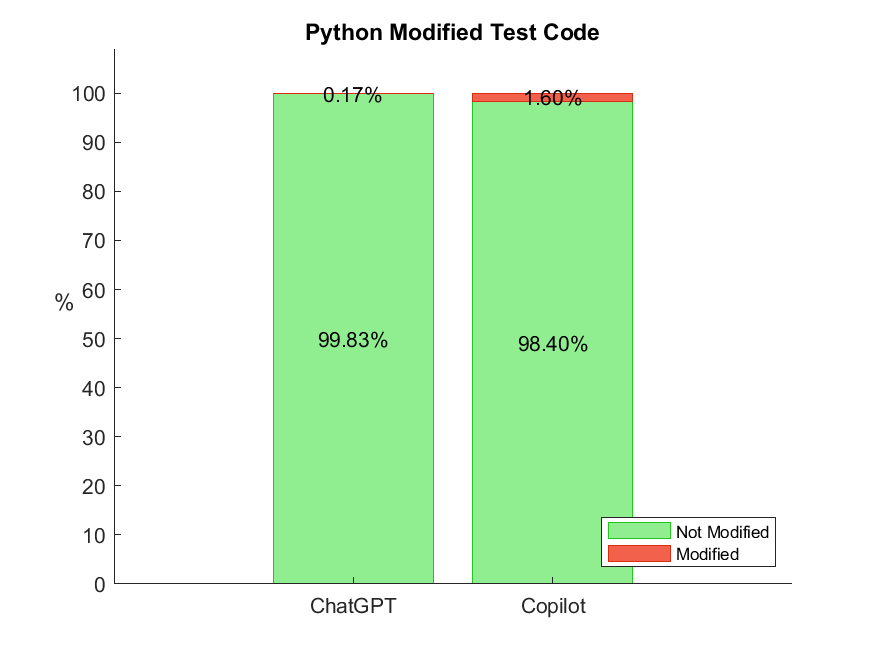}
        \captionsetup{justification=centering}
        \caption{Modification rate results for Python}
        \label{fig:scg-modrate-python}
    \end{minipage}
\end{figure}

As discussed in Section~\ref{sec:manCorrection}, we needed to manually modify some unit test cases in order to compile these files and receive a report on the coverage. Evaluating these required modifications demonstrates that only minor adjustments to the code samples were needed. This result applies to both Java and Python, see Figures~\ref{fig:scg-modrate-java} and~\ref{fig:scg-modrate-python}.

In general, we made very few modifications. Comparing ChatGPT and GitHub Copilot, there is hardly any noticeable difference between them for the two languages. However, this difference between ChatGPT and Copilot is still statistically significant at a significance level of 0.01, as demonstrated by a Mann-Whitney U test performed separately for Java and Python.

No difference between the two programming languages is recognizable, a fact which is further supported by a Mann Whitney U test based on a significance level of 0.05.

\subsection{Differences Over Time}

We investigate differences in the outcomes over time by rerunning the experiment of the baseline study~\cite{HE2023} and generating code for the identical six Java algorithms half a year later. Our objective was to identify noticeable and significant changes in the correctness and quality of the code.

The difference over time in ChatGPT's performance is substantial, while Copilot's correctness is almost similar. Comparing our results with theirs for both AI models in the six algorithms of the baseline study using the Mann-Whitney U test, we identify a significant difference in code correctness for ChatGPT at the 0.05 significance level, while the difference is not significant for GitHub Copilot. 

We also identify variations in code quality. The baseline study attained a 98.52~\% code quality using ChatGPT and 94.07~\% with GitHub Copilot. While ChatGPT achieved a similar performance with a 98~\% rate of quality compliant generated lines in our study, GitHub Copilot improved its performance by 4~\% to 98.05~\%. The statistical analysis confirms a significant difference in the code quality of GitHub Copilot between our results and the baseline study at the 0.05 level of significance. However, ChatGPT did not show any significant improvement. To summarize these tests, there is an improvement of AI-generated code over time. Specifically, ChatGPT excels in ensuring code correctness, while GitHub Copilot enhanced its code quality to match that of ChatGPT.

\begin{figure}[ht]
    \centering
    \begin{minipage}[c]{.49\textwidth}
        \centering
        \includegraphics[width=1\linewidth]{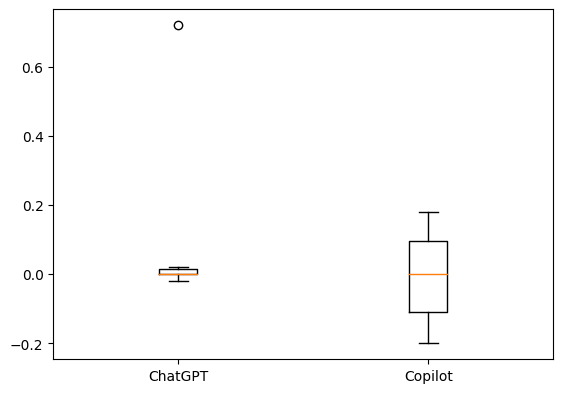}
        \captionsetup{justification=centering}
        \caption{Correctness changes between $t_0$ and $t_1$}
        \label{fig:scg-change-correctness}
    \end{minipage}
    \begin{minipage}[c]{.49\textwidth}
        \centering
        \includegraphics[width=1\linewidth]{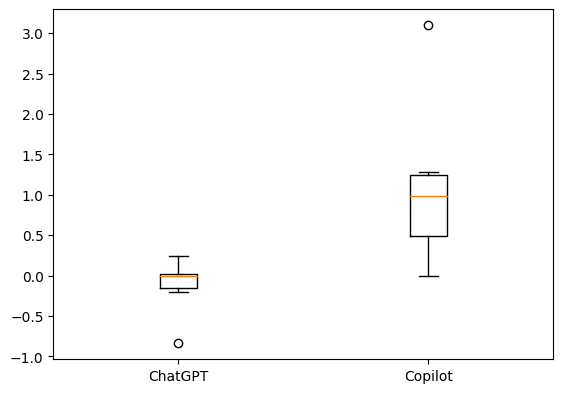}
        \captionsetup{justification=centering}
        \caption{Quality changes between $t_0$ and $t_1$}
        \label{fig:scg-change-quality}
    \end{minipage}
\end{figure}

Figures~\ref{fig:scg-change-correctness} and~\ref{fig:scg-change-quality} show the correctness and quality differences over time, i.e., the differences of the means of 50 generations for each algorithm generated at $t_0$ and $t_1$, respectively. Here, only quality score differences for Copilot are noticeably larger than zero, indicating a noticeable improvement from $t_0$ to $t_1$: on average, the number of quality issues has dropped by $\approx 1$. However, according to the Sign test this difference is not significant at the 0.05 significance level. In contrast to that, the less conservative Wilcoxon signed-rank test does indicate a significant improvement at this level. In the three other cases (correctness changes for both LLMs and quality changes for ChatGPT), both tests fail in rejecting the null hypothesis that the (barely noticeable) differences are just random at the 0.05 significance level. Mind the outlier algorithm in the correctness measurement of ChatGPT indicating a 70\% improvement from $t_0$ to $t_1$. This outlier was disregarded in the paired difference tests, which explains the  conclusion differing from the Mann-Whitney U test results. 

\section{Discussion}\label{sec:discussion}

\subsection{Analysis Results}
We can summarize that ChatGPT does very well in generating correct code. ChatGPT achieved 89.33~\% for Java code and 79.17~\% for Python code in terms of code correctness. On the other hand, GitHub Copilot does also manage to generate correct code. It reached 75.50~\% for Java code and 62.50~\% for Python code in terms of code correctness. These results answer the first research question: \textit{1. How well does ChatGPT (GitHub Copilot) provide correct code in Python (Java) from specified instructions?} However, there are threads to validity, as discussed in Section~\ref{sec:threats}.

The results from ChatGPT in code quality were also good, with 98.09~\%\ for Java code and 88.20~\% for Python code. In contrast to the code correctness of code generated by GitHub Copilot, the code quality of the GitHub Copilot generated code showed better results. It achieved 98.13~\% for Java code and 90.15~\% for Python code. These results answer the second research question: \textit{2. How good is ChatGPT (GitHub Copilot) at generating Python (Java) code of high quality?}

For test case generation, ChatGPT achieves only intermediate results in code correctness. The outcomes were 37.50~\% on average for the three approaches for Java code and 28.61~\% on average for the three approaches for Python code. GitHub Copilot performed slightly better and achieved 49.72~\% on average for the three approaches for Java code and 39.17~\% on average for the three approaches for Python code in terms of code correctness. Only little manual work, about 1~\%, was needed to correct the generated test cases. This answers the third research questions \textit{3. How well does ChatGPT (GitHub Copilot) provide correct Unit Tests in Python (Java) from specified instructions?}

Regarding test code quality, ChatGPT shows intermediate coverage in Java and good coverage in Python, i.e., 58.79~\% on average for the three approaches for Java code and 85.94~\% on average for the three approaches for Python code. GitHub Copilot similarly in coverage; the code coverage was 57.18~\% on average for the three approaches for Java code and 82.03~\% on average for the three approaches for Python code. This answers the third research questions \textit{4. How good is ChatGPT (GitHub Copilot) at generating Python (Java) Unit Tests of high quality?}

Comparing Java and Python shows, that both AI models achieved better results in Java than in Python with a statistically significant difference based on a significance level of 0.01 in terms of code correctness and quality. For test case generation, Java also performs better in test correctness. However, Python shows better results in test coverage, with reaching nearly total coverage. There is a statistically significant difference in the test correctness and coverage based on a significance level of 0.01 between the languages Java and Python. These statistically significant differences were determined using the Mann-Whitney U test. This answers the fifth research question \textit{5. Can we observe significant differences between Java and Python?} 

All following statistically significant differences were determined using the Mann-Whitney U test.
There is a statistically significant difference between ChatGPT and GitHub Copilot:
\begin{itemize}
    \item in the code correctness of generated Java and Python codes, resp. ($p < 0.01$),
    \item in the code quality of generated Java and Python codes, resp. ($p < 0.05$),
    \item in the test code correctness of generated Java and Python codes, resp. ($p < 0.01$),
    \item in the test code coverage of generated Java code ($p < 0.01$).
\end{itemize} 
Given these differences in the performance between the two AI models, ChatGPT seems to be the choice for generating correct, high-quality code and test cases. On the other hand, GitHub Copilot is not far away from the performance of ChatGPT, noting also the convenience to use this tool directly in the developer's IDE. This provides an answer to the research question \textit{6. Can we observe significant differences between ChatGPT and GitHub Copilot?} 

There is a  improvement of LLM-generated code over time. ChatGPT excels in code correctness and GitHub Copilot in code quality. The former is supported only by the Mann-Whitney U test (assuming independent samples), the latter even by the Wilcoxon signed-rank test (assuming dependent samples). This answers the research question \textit{7. Can we observe significant changes compared to the baseline study~\cite{HE2023}, differences over time?} 

\subsection{Threats to Validity}\label{sec:threats}

In this section, we discuss internal and external threats to the validity of our study. Internal validity refers to the degree of confidence that the causal relationships being tested in this study are trustworthy and not influenced by other factors or variables. External validity refers to the extent to which results from this study can be applied and generalized to other situations, groups, or events.

\subsubsection{Internal Validity}

Due to the non-deterministic nature of the LLMs, the results of one and the same prompt are not repeatable to 100~\%. We mitigate this problem by generating several implementations resetting the session context in between but differences remain. Although the differences appear to be minor, an analysis of the statistical significance of these differences would be one of the possibilities to extend this study, as explained in Section~\ref{sec:conclude}.

Another threat to the validity of the above results was already observed during the course of our experiment. We run the generations in parallel using devices with different specifications and operating systems. Due to the nondeterminism in some assessment tools and even in some of the generated algorithms, results were not the same for different users and machines, even though the data generated was the same. Even after using the same versions of each tool involved in the examination process, the results on different devices still differed from time to time. We did our best to find the reason for these differences and correct them where possible, but we were unable to correct them all. Again the remaining differences are small, but an analysis of statistical significance of these differences would be a possible extension of this study, see Section~\ref{sec:conclude}.

Moreover, the correction of unusable test case generations required human intervention. This might lead to different approaches to adapting the code and therefore different results. The potentiality of executing the experiment differently from the intention of the instructions is also a threat to the equivalence of our results and to future repetitions of our experiment.

Finally, the benchmark code was derived from standard algorithms; similar implementations might have leaked into the training set of the LLMs, cf. the discussion in Section \ref{sec:related}.

\textbf{External Validity}
Since GitHub Copilot receives continuous updates without a specific version available for selection, the possibility of future variations in results is more likely than not. This is due to the ongoing development of the AI model, indicating that outcomes from GitHub Copilot's future generations may be superior or inferior to the ones obtained in our experiment. A similar issue applies to ChatGPT. It is to be expected that the GPT LLM will undergo further improvements, and therefore, the possibility of new versions of the model generating better or worse results. OpenAI might even choose to eliminate GPT-3.5 as a public version, making our results harder to repeat. 

Moreover, the study centered on Java and Python, both of which rank among the top ten most frequently used programming languages \cite{most_used_pl}. Nonetheless, it is possible that these two languages do not provide a fair representation of the vast number of programming languages that exist. 

Likewise, the algorithms selected primarily by following the list of ``25 algorithms every programmer should know'' may be rather simple compared to other algorithms, which may have had a positive effect on the results of the study. The study should be extended with more and less popular algorithm to generalize the results. 

Also, we chose English as the language of the prompts, since we expected the most promising results there. The use of other languages may result in different results, as well. 

We created the prompts at our own discretion and not according to a specific, fixed pattern nor a with a maximum length. Although we aimed for short and concise prompts, remaining differences may have influenced the results: more precisely specified prompts may produce better results than simpler and shorter prompts.

Finally, the benchmark code was derived from standard algorithms, an approach that has been criticized for lacking practical relevance, cf. Section \ref{sec:related}.

\section{Related Work}\label{sec:related}

E. Hansson and O. Ellréus evaluated the code generation abilities of GitHub Copilot and ChatGPT in terms of code correctness and code quality in early 2023~\cite{HE2023}. On a subset of our algorithms, their study shows that both ChatGPT and GitHub Copilot are capable of producing Java code given specific instructions. ChatGPT had attained a correctness rate of 87~\%, while GitHub Copilot achieved a rate of 89~\%. Half a year later and with an extended set of algorithms, we showed that ChatGPT increased the correctness rate to 89~\%, while GitHub Copilot's correctness rate dropped to 75.5~\%. Furthermore, their study could not show a significant difference in code correctness between the two tools, while our study did. Moreover, in early 2023 ChatGPT managed to produce 98.52~\% of its generated lines without quality rule violations, whereas GitHub Copilot has produced only 94.07~\% of its generated lines without any quality rule violations. In terms of code quality, their analysis shows that ChatGPT is statistically significantly better than GitHub Copilot. In our study, both tools generated code with more than 98~\% of its lines without quality rule violations; a significant difference of the tools couldn't be observed any longer.

In addition to the study of Hansson and  Ellréus~\cite{HE2023}, ours incorporated six additional algorithms in Java. Moreover, we assessed the code correctness and code quality for the corresponding algorithms in Python as well. Furthermore, we investigated the ability of both tools to generate unit tests in the mentioned programming languages. 

Chen et al. conducted a study assessing the functional correctness of code produced by different LLMs \cite{Chenetal2021}. The study examined GPT-3/-J's code generation abilities, with particular emphasis on the Codex model, which is fine-tuned on publicly available code from GitHub. It was found that Codex is capable of solving 28.8~\% of problems in the HumanEval dataset, whereas GPT3 solves none and GPT-J solves 11.4~\%. The research concentrated on generating isolated Python functions from Docstrings (documentation string literals after the function definition). 

There exists also a variety of alternative Python benchmarks including HumanEval \cite{Chenetal2021}, MBPP~\cite{Austinetal21}, APPS~\cite{Hindleetal16}, and DS-1000~\cite{Laietal23}. They are larger than our benchmark ranging from a few hundred instances in HumanEval to several thousand instances in APPS and MBPP. Yadav et al. showed the benchmarks bias towards a limited set of programming concepts, neglecting most of the other concepts entirely, and suggests PythonSaga~\cite{Yadavetal24}. All these benchmarks are limited to Python. Therefore, Zheng et al. introduced HumanEval-X, a benchmark for evaluating multilingual models in Python, C++, Java, JavaScript, and Go~\cite{Zhengetal23}. For each problem defined in Humelval only for Python, they manually rewrote its prompt, canonical solution, and test cases in the other four languages. Dou et al. 
criticize that benchmarks commonly used are constructed from programming questions, hence, disconnected from real-world programming, or derived from real-world code repositories and, hence, contaminated as also used for pre-training the LLMs~\cite{douetal24}. Riddell et al. even quantifies this contamination~\cite{riddelletal24}. We developed our own benchmark from programming questions admitting that relevance and contamination are hard to assess; both are threads to validity. 

Using different Python benchmarks, Yeo et al. suggest a framework for the evaluation of LLM-generated code including metrics ($\mathit{pass}@n$) for the fine grained assessment of the results~\cite{Yeoetal24}. Zheng et al. evaluated their model CodeGeeX using HumanEval-X and ,
the ($\mathit{pass}@n$) metric for Python, C++, Java, JavaScript, and Go~\cite{Zhengetal23}. However, they do not suggest the generation and evaluation of test codes and their assessment as the present paper does. Repeating our tests with (the Python and Java problems of) HumanEval-X is matter of future work.

Evaluating the ability of LLMs to generate correct and high-quality code was also investigated by many other researchers. Here some examples: Niu et al. evaluated the code efficiency \cite{Niuetal24} using the framework suggested in \cite{Yeoetal24}. Yetistiren et al. evaluated  code generation of GitHub Copilot in terms of correctness, validity, and efficiency \cite{YOT2022}. They discovered that GitHub Copilot could create 28.7~\% fully correct code for problems in the HumanEval dataset. For this dataset, 51.2~\% of generated code was partially correct, whereas 20.1~\% was incorrect. 

Wang et al. found inadequate quality characteristics considered in the existing studies about code generated by LLMs, if quality was considered at all~\cite{Wangetal23}. We evaluated and compared the code quality of ChatGPT and GitHub Copilot with well-established Clean Code~\cite{Martin2008} quality characteristics.

Comparably few researchers evaluated the ability of LLMs to generate correct and high-quality {\em test} code: Li et al. and Chen et al. show that the $\mathit{pass}@n$ metric can be improved by generating tests using LLMs in their systems AlphaCode~\cite{lietal22} and CodeT~\cite{chenetal22}, resp. Lahiri, Fakhoury et al. evaluated the development efficiency, correctness, and cognitive load of (test) code in their test-driven development workflow~\cite{lahirietal23,fakhouryetal24}. None of these papers on test code generation evaluated the coverage of the generated tests.

Pan et al. addressed a related problem: they studied the performance of converting source code from one programming language to another across pairs of different languages, including C, C++, Go, Java, and Python~\cite{Panetal23}. Their prompts included of the source code implementation of a method, not only a description of the method as in our case, but they propose a prompt-crafting approach to provide additional context for LLMs. In addition to the distinction of syntactical and functional errors, as we did in our study, they also identify root causes for translation bugs. We aimed to evaluate the coding capabilities of ChatGPT and GitHub Copilot using a prompt text as a comprehensive comment, rather than using Docstrings. We also intended to insert code snippets within the introductory section to assess tool performance in this scenario.

\section{Conclusion and Future Work}\label{sec:conclude}

The study indicates that LLMs can generate algorithms and test codes for both Java and Python. Algorithm generation works, however, way better than test case generation. It can also be stated that (currently) ChatGPT  outperforms GitHub Copilot in generating correct code, although both yield code of excellent quality. 
When producing test cases, both AI models produce similar results in terms of test coverage, with ChatGPT performing better on some (prompting) approaches. However, in terms of the correctness of generated test cases, GitHub Copilot produces more correct outcomes than ChatGPT. Regardless of the LLM, the code generated in Java achieved significantly better results overall than the code generated for Python, with test coverage results the only exception.

However, given the shortfall of 10~\% left to attain faultless code in rather well-known standard algorithms, it is reasonable to assume that the era of artificial intelligence superseding human software developers is yet to come. Nevertheless, both ChatGPT and GitHub Copilot prove to be useful assisting tools in developing software, enabling the developers to focus on more complex tasks in their job. 

Further research is required to provide a more elaborate conclusion on the methods and opportunities available for supporting and automating software development, as discussed in Section~\ref{sec:discussion}. This future work involves categorizing the AI tasks according to their complexity, which may reveal potential effects on correctness and quality of the generated (test) codes. Since the research on the code generation capabilities of LLMs is an ongoing topic, one could compare the results of this and other studies to try to identify a trend in code correctness and code quality over time. Another way to increase the variety of the code generation tasks, is to assess the capabilities of LLMs to refactor or debug already written code. Finally, it is interesting to statistically analyze the influence of non-determinism in (test) code generation, in the generated algorithms, and experimental environment on the study results.

\bibliography{main}

\begin{thebibliography}{39}
\providecommand{\natexlab}[1]{#1}
\providecommand{\url}[1]{\texttt{#1}}
\providecommand{\urlprefix}{}

\bibitem[{GitHub(2023)}]{TopLang2023}
GitHub, The top programming languages; 2023.
\newblock Accessed for describing this study in November 2023.
\newblock \url{https://octoverse.github.com/2022/top-programming-languages}.

\bibitem[{OpenAI(2023)}]{OpenAI2023}
OpenAI, ChatGPT; 2023.
\newblock Accessed for this study in September 2023.
\newblock \url{https://chat.openai.com/chat}.

\bibitem[{GitHub(2023)}]{Github2023}
GitHub, Copilot; 2023.
\newblock Accessed for this study in September 2023.
\newblock \url{https://github.com/features/copilot}.

\bibitem[{AI(2023)Unite AI}]{CGtools2023}
AI U, 10 Best AI Code Generators; 2023.
\newblock Accessed for describing this study in November 2023.
\newblock \url{https://www.unite.ai/best-ai-code-generators/}.

\bibitem[{Du et~al.(2023)Du, Mengnan and He, Fengxiang and Zou, Na and Tao,
  Dacheng and Hu, Xia}]{du_shortcut_2023}
Du M, He F, Zou N, Tao D, Hu X, Shortcut {Learning} of {Large} {Language}
  {Models} in {Natural} {Language} {Understanding}.
\newblock arXiv; 2023.
\newblock \urlprefix\url{http://arxiv.org/abs/2208.11857}, number:
  arXiv:2208.11857 arXiv:2208.11857 [cs].

\bibitem[{Chang et~al.(2023)Chang, Yupeng and Wang, Xu and Wang, Jindong and
  Wu, Yuan and Yang, Linyi and Zhu, Kaijie and Chen, Hao and Yi, Xiaoyuan and
  Wang, Cunxiang and Wang, Yidong and Ye, Wei and Zhang, Yue and Chang, Yi and
  Yu, Philip S. and Yang, Qiang and Xie, Xing}]{chang_survey_2023}
Chang Y, Wang X, Wang J, Wu Y, Yang L, Zhu K, et~al., A {Survey} on
  {Evaluation} of {Large} {Language} {Models}.
\newblock arXiv; 2023.
\newblock \urlprefix\url{http://arxiv.org/abs/2307.03109}, number:
  arXiv:2307.03109 arXiv:2307.03109 [cs].

\bibitem[{Douglas(2023)Douglas, Michael R.}]{douglas_large_2023}
Douglas MR, Large {Language} {Models}.
\newblock arXiv; 2023.
\newblock \urlprefix\url{http://arxiv.org/abs/2307.05782}, number:
  arXiv:2307.05782 arXiv:2307.05782 [hep-th, physics:physics].

\bibitem[{cha(????)}]{chatgpt_explanation}
What is {ChatGPT}? {\textbar} {OpenAI} {Help} {Center};.
\newblock
  \urlprefix\url{https://help.openai.com/en/articles/6783457-what-is-chatgpt}.

\bibitem[{Vaswani et~al.(2017)Vaswani, Ashish and Shazeer, Noam and Parmar,
  Niki and Uszkoreit, Jakob and Jones, Llion and Gomez, Aidan N and Kaiser, \L
  ukasz and Polosukhin, Illia}]{Vaswanietal2017}
Vaswani A, Shazeer N, Parmar N, Uszkoreit J, Jones L, Gomez AN, et~al.
\newblock Attention is All you Need.
\newblock In: Guyon I, Luxburg UV, Bengio S, Wallach H, Fergus R, Vishwanathan
  S, et~al., editors. Advances in Neural Information Processing Systems,
  vol.~30 Curran Associates, Inc.; 2017.
  \urlprefix\url{https://proceedings.neurips.cc/paper_files/paper/2017/file/3f5ee243547dee91fbd053c1c4a845aa-Paper.pdf}.

\bibitem[{Radford et~al.(????)Radford, Alec and Wu, Jeffrey and Child, Rewon
  and Luan, David and Amodei, Dario and Sutskever, Ilya}]{lm_unsupervised}
Radford A, Wu J, Child R, Luan D, Amodei D, Sutskever I, Language {Models} are
  {Unsupervised} {Multitask} {Learners};.

\bibitem[{Brown et~al.(2020)Brown, Tom B. and Mann, Benjamin and Ryder, Nick
  and Subbiah, Melanie and Kaplan, Jared and Dhariwal, Prafulla and
  Neelakantan, Arvind and Shyam, Pranav and Sastry, Girish and Askell, Amanda
  and Agarwal, Sandhini and Herbert-Voss, Ariel and Krueger, Gretchen and
  Henighan, Tom and Child, Rewon and Ramesh, Aditya and Ziegler, Daniel M. and
  Wu, Jeffrey and Winter, Clemens and Hesse, Christopher and Chen, Mark and
  Sigler, Eric and Litwin, Mateusz and Gray, Scott and Chess, Benjamin and
  Clark, Jack and Berner, Christopher and McCandlish, Sam and Radford, Alec and
  Sutskever, Ilya and Amodei, Dario}]{lm_few_shot}
Brown TB, Mann B, Ryder N, Subbiah M, Kaplan J, Dhariwal P, et~al., Language
  {Models} are {Few}-{Shot} {Learners}.
\newblock arXiv; 2020.
\newblock \urlprefix\url{http://arxiv.org/abs/2005.14165}, arXiv:2005.14165
  [cs].

\bibitem[{Dale(2021)Dale, Robert}]{gpt-3}
Dale R.
\newblock {GPT}-3: {What}’s it good for?
\newblock Natural Language Engineering 2021 Jan;27(1):113--118.
\newblock
  \urlprefix\url{https://www.cambridge.org/core/journals/natural-language-engineering/article/gpt3-whats-it-good-for/0E05CFE68A7AC8BF794C8ECBE28AA990},
  publisher: Cambridge University Press.

\bibitem[{cha(????)}]{chatgpt-4_payment}
How can {I} use {GPT}-4 in {ChatGPT}? {\textbar} {OpenAI} {Help} {Center};.
\newblock
  \urlprefix\url{https://help.openai.com/en/articles/7127997-how-can-i-use-gpt-4-in-chatgpt}.

\bibitem[{cop(????)}]{copilot_explanation}
About {GitHub} {Copilot} for {Individuals};.
\newblock
  \urlprefix\url{https://ghdocs-prod.azurewebsites.net/en/copilot/overview-of-github-copilot/about-github-copilot-for-individuals}.

\bibitem[{Chen et~al.(2021)Mark Chen and Jerry Tworek and Heewoo Jun and Qiming
  Yuan and Henrique Ponde and Jared Kaplan and Harrison Edwards and Yura Burda
  and Nicholas Joseph and Greg Brockman and Alex Ray and Raul Puri and Gretchen
  Krueger and Michael Petrov and Heidy Khlaaf and Girish Sastry and Pamela
  Mishkin and Brooke Chan and Scott Gray and Nick Ryder and Mikhail Pavlov and
  Alethea Power and Lukasz Kaiser and Mohammad Bavarian and Clemens Winter and
  Philippe Tillet and Felipe Petroski Such and David W. Cummings and Matthias
  Plappert and Fotios Chantzis and Elizabeth Barnes and Ariel Herbert-Voss and
  William H. Guss and Alex Nichol and Igor Babuschkin and S. Arun Balaji and
  Shantanu Jain and Andrew Carr and Jan Leike and Joshua Achiam and Vedant
  Misra and Evan Morikawa and Alec Radford and Matthew M. Knight and Miles
  Brundage and Mira Murati and Katie Mayer and Peter Welinder and Bob McGrew
  and Dario Amodei and Sam McCandlish and Ilya Sutskever and Wojciech
  Zaremba}]{Chenetal2021}
Chen M, Tworek J, Jun H, Yuan Q, Ponde H, Kaplan J, et~al.
\newblock Evaluating Large Language Models Trained on Code.
\newblock ArXiv 2021;abs/2107.03374.
\newblock \urlprefix\url{https://api.semanticscholar.org/CorpusID:235755472}.

\bibitem[{Hansson and Ellréus(2023)Hansson, Emilia and Ellréus,
  Oliwer}]{HE2023}
Hansson E, Ellréus O, Code Correctness and Quality in the Era of AI Code
  Generation: Examining ChatGPT and GitHub Copilot; 2023.

\bibitem[{Cutajar(2018)James Cutajar}]{Cutajar2018}
Cutajar J.
\newblock Beginning Java Data Structures and Algorithms: Sharpen your problem
  solving skills by learning core computer science concepts in a pain-free
  manner.
\newblock Birmingham, Mumbai: Packt Publishing; 2018.

\bibitem[{Srivastava(2023)Vivek Srivastava}]{Srivastava2023}
Srivastava V, Top 25 Algorithms Every Programmer Should Know; 2023.
\newblock Accessed 9/2023.
\newblock
  \url{https://medium.com/techie-delight/top-25-algorithms-every-programmer-should-know-373246b4881b}.

\bibitem[{Martin(2008)Martin, Robert C.}]{Martin2008}
Martin RC.
\newblock Clean Code: A Handbook of Agile Software Craftsmanship.
\newblock USA: Pearson Education; 2008.

\bibitem[{Van~Rossum et~al.(2001)Van Rossum, Guido and Warsaw, Barry and
  Coghlan, Alyssa}]{PEP2001}
Van~Rossum G, Warsaw B, Coghlan A, PEP 8 – Style Guide for Python Code; 2001.
\newblock Accessed: 29.10.2023.
\newblock \url{https://peps.python.org/pep-0008/}.

\bibitem[{Levenshtein(1966)Vladimir I. Levenshtein}]{Levenshtein1966}
Levenshtein VI.
\newblock Binary codes capable of correcting deletions, insertions, and
  reversals, vol.~10; 1966.

\bibitem[{Verma and Abdel-Salam(2019)Verma, J. P. and Abdel-Salam, Abdel-Salam
  G.}]{VA2019}
Verma JP, Abdel-Salam ASG.
\newblock Testing Statistical Assumptions in Research.
\newblock USA: John Wiley \& Sons, Inc.; 2019.

\bibitem[{mos(????)}]{most_used_pl}
Most used languages among software developers globally 2023;.
\newblock
  \urlprefix\url{https://www.statista.com/statistics/793628/worldwide-developer-survey-most-used-languages/}.

\bibitem[{Austin et~al.(2021)Jacob Austin and Augustus Odena and Maxwell I. Nye
  and Maarten Bosma and Henryk Michalewski and David Dohan and Ellen Jiang and
  Carrie J. Cai and Michael Terry and Quoc V. Le and Charles
  Sutton}]{Austinetal21}
Austin J, Odena A, Nye MI, Bosma M, Michalewski H, Dohan D, et~al.
\newblock Program Synthesis with Large Language Models.
\newblock CoRR 2021;abs/2108.07732.
\newblock \urlprefix\url{https://arxiv.org/abs/2108.07732}.

\bibitem[{Hindle et~al.(2016)Hindle, Abram and Barr, Earl T. and Gabel, Mark
  and Su, Zhendong and Devanbu, Premkumar}]{Hindleetal16}
Hindle A, Barr ET, Gabel M, Su Z, Devanbu P.
\newblock On the naturalness of software.
\newblock Commun ACM 2016 apr;59(5):122–131.
\newblock \urlprefix\url{https://doi.org/10.1145/2902362}.

\bibitem[{Lai et~al.(2023)Yuhang Lai and Chengxi Li and Yiming Wang and Tianyi
  Zhang and Ruiqi Zhong and Luke Zettlemoyer and Wen-tau Yih and Daniel Fried
  and Sida I. Wang and Tao Yu}]{Laietal23}
Lai Y, Li C, Wang Y, Zhang T, Zhong R, Zettlemoyer L, et~al.
\newblock DS-1000: A Natural and Reliable Benchmark for Data Science Code
  Generation.
\newblock In: 0001 AK, Brunskill E, Cho K, Engelhardt B, Sabato S, Scarlett J,
  editors. International Conference on Machine Learning, ICML 2023, 23-29 July
  2023, Honolulu, Hawaii, USA, vol. 202 of Proceedings of Machine Learning
  Research PMLR; 2023. p. 18319--18345.
\newblock \urlprefix\url{https://proceedings.mlr.press/v202/lai23b.html}.

\bibitem[{Yadav et~al.(2024)Ankit Yadav and Himanshu Beniwal and Mayank
  Singh}]{Yadavetal24}
Yadav A, Beniwal H, Singh M.
\newblock PythonSaga: Redefining the Benchmark to Evaluate Code Generating
  LLMs; 2024.
  \urlprefix\url{https://api.semanticscholar.org/CorpusID:266844287}.

\bibitem[{Zheng et~al.(2023)Zheng, Qinkai and Xia, Xiao and Zou, Xu and Dong,
  Yuxiao and Wang, Shan and Xue, Yufei and Shen, Lei and Wang, Zihan and Wang,
  Andi and Li, Yang and Su, Teng and Yang, Zhilin and Tang, Jie}]{Zhengetal23}
Zheng Q, Xia X, Zou X, Dong Y, Wang S, Xue Y, et~al.
\newblock CodeGeeX: A Pre-Trained Model for Code Generation with Multilingual
  Benchmarking on HumanEval-X.
\newblock In: Proceedings of the 29th ACM SIGKDD Conference on Knowledge
  Discovery and Data Mining KDD '23, New York, NY, USA: Association for
  Computing Machinery; 2023. p. 5673–5684.
\newblock \urlprefix\url{https://doi.org/10.1145/3580305.3599790}.

\bibitem[{Dou et~al.(2024)Shihan Dou and Haoxiang Jia and Shenxi Wu and Huiyuan
  Zheng and Weikang Zhou and Muling Wu and Mingxu Chai and Jessica Fan and
  Caishuang Huang and Yunbo Tao and Yan Liu and Enyu Zhou and Ming Zhang and
  Yuhao Zhou and Yueming Wu and Rui Zheng and Ming Wen and Rongxiang Weng and
  Jingang Wang and Xunliang Cai and Tao Gui and Xipeng Qiu and Qi Zhang and
  Xuanjing Huang}]{douetal24}
Dou S, Jia H, Wu S, Zheng H, Zhou W, Wu M, et~al., What's Wrong with Your Code
  Generated by Large Language Models? An Extensive Study; 2024.
\newblock \urlprefix\url{https://arxiv.org/abs/2407.06153}.

\bibitem[{Riddell et~al.(2024)Riddell, Martin and Ni, Ansong and Cohan,
  Arman}]{riddelletal24}
Riddell M, Ni A, Cohan A.
\newblock Quantifying Contamination in Evaluating Code Generation Capabilities
  of Language Models.
\newblock In: Ku LW, Martins A, Srikumar V, editors. Proceedings of the 62nd
  Annual Meeting of the Association for Computational Linguistics (Volume 1:
  Long Papers) Bangkok, Thailand: Association for Computational Linguistics;
  2024. p. 14116--14137.
\newblock \urlprefix\url{https://aclanthology.org/2024.acl-long.761}.

\bibitem[{Yeo et~al.(2024)Yeo, Sangyeop and Ma, Yu-Seung and Kim, Sang Cheol
  and Jun, Hyungkook and Kim, Taeho}]{Yeoetal24}
Yeo S, Ma YS, Kim SC, Jun H, Kim T.
\newblock Framework for evaluating code generation ability of large language
  models.
\newblock ETRI Journal 2024;46(1):106--117.
\newblock
  \urlprefix\url{https://onlinelibrary.wiley.com/doi/abs/10.4218/etrij.2023-0357}.

\bibitem[{Niu et~al.(2024)Niu, Changan and Zhang, Ting and Li, Chuanyi and Luo,
  Bin and Ng, Vincent}]{Niuetal24}
Niu C, Zhang T, Li C, Luo B, Ng V.
\newblock On Evaluating the Efficiency of Source Code Generated by LLMs.
\newblock In: Proceedings of the 2024 IEEE/ACM First International Conference
  on AI Foundation Models and Software Engineering FORGE '24, New York, NY,
  USA: Association for Computing Machinery; 2024. p. 103–107.
\newblock \urlprefix\url{https://doi.org/10.1145/3650105.3652295}.

\bibitem[{Yetistiren et~al.(2022)Yetistiren, Burak and Ozsoy, Isik and Tuzun,
  Eray}]{YOT2022}
Yetistiren B, Ozsoy I, Tuzun E.
\newblock Assessing the Quality of GitHub Copilot’s Code Generation.
\newblock In: Proceedings of the 18th International Conference on Predictive
  Models and Data Analytics in Software Engineering PROMISE 2022, New York, NY,
  USA: Association for Computing Machinery; 2022. p. 62–71.
\newblock \urlprefix\url{https://doi.org/10.1145/3558489.3559072}.

\bibitem[{Wang and Chen(2023)Wang, Jianxun and Chen, Yixiang}]{Wangetal23}
Wang J, Chen Y.
\newblock A Review on Code Generation with LLMs: Application and Evaluation.
\newblock In: 2023 IEEE International Conference on Medical Artificial
  Intelligence (MedAI); 2023. p. 284--289.

\bibitem[{Li et~al.(2022)Li, Yujia and Choi, David and Chung, Junyoung and
  Kushman, Nate and Schrittwieser, Julian and Leblond, Rémi and Eccles, Tom
  and Keeling, James and Gimeno, Felix and Dal Lago, Agustin and Hubert, Thomas
  and Choy, Peter and de Masson d’Autume, Cyprien and Babuschkin, Igor and
  Chen, Xinyun and Huang, Po-Sen and Welbl, Johannes and Gowal, Sven and
  Cherepanov, Alexey and Molloy, James and Mankowitz, Daniel J. and Sutherland
  Robson, Esme and Kohli, Pushmeet and de Freitas, Nando and Kavukcuoglu, Koray
  and Vinyals, Oriol}]{lietal22}
Li Y, Choi D, Chung J, Kushman N, Schrittwieser J, Leblond R, et~al.
\newblock Competition-level code generation with AlphaCode.
\newblock Science 2022 Dec;378(6624):1092–1097.
\newblock \urlprefix\url{http://dx.doi.org/10.1126/science.abq1158}.

\bibitem[{Chen et~al.(2022)Bei Chen and Fengji Zhang and Anh Nguyen and
  Daoguang Zan and Zeqi Lin and Jian-Guang Lou and Weizhu Chen}]{chenetal22}
Chen B, Zhang F, Nguyen A, Zan D, Lin Z, Lou JG, et~al., CodeT: Code Generation
  with Generated Tests; 2022.
\newblock \urlprefix\url{https://arxiv.org/abs/2207.10397}.

\bibitem[{Lahiri et~al.(2023)Shuvendu K. Lahiri and Sarah Fakhoury and Aaditya
  Naik and Georgios Sakkas and Saikat Chakraborty and Madanlal Musuvathi and
  Piali Choudhury and Curtis von Veh and Jeevana Priya Inala and Chenglong Wang
  and Jianfeng Gao}]{lahirietal23}
Lahiri SK, Fakhoury S, Naik A, Sakkas G, Chakraborty S, Musuvathi M, et~al.,
  Interactive Code Generation via Test-Driven User-Intent Formalization; 2023.
\newblock \urlprefix\url{https://arxiv.org/abs/2208.05950}.

\bibitem[{Fakhoury et~al.(2024)Sarah Fakhoury and Aaditya Naik and Georgios
  Sakkas and Saikat Chakraborty and Shuvendu K. Lahiri}]{fakhouryetal24}
Fakhoury S, Naik A, Sakkas G, Chakraborty S, Lahiri SK, LLM-based Test-driven
  Interactive Code Generation: User Study and Empirical Evaluation; 2024.
\newblock \urlprefix\url{https://arxiv.org/abs/2404.10100}.

\bibitem[{Pan et~al.(2023)Rangeet Pan and Ali Reza Ibrahimzada and Rahul
  Krishna and Divya Sankar and Lambert Pouguem Wassi and Michele Merler and
  Boris Sobolev and Raju Pavuluri and Saurabh Sinha and Reyhaneh
  Jabbarvand}]{Panetal23}
Pan R, Ibrahimzada AR, Krishna R, Sankar D, Wassi LP, Merler M, et~al.,
  Understanding the Effectiveness of Large Language Models in Code Translation;
  2023.

\end{thebibliography}

\appendix

\section{Additional Information} \label{app:repos}
Further information regarding the study is available in our GitHub repositories, which were established to perform the experiment. The Source Code repository incorporates ChatGPT's and GitHub Copilot's generated code and unit tests in Java and Python, along with pipeline scripts for automated assessment and guidance on how to re-run the experiment. The Analysis Repository presents the evaluation and analysis of the results, with scripts to generate the graphs and all the raw data. Source Code Repository:  \url{https://github.com/tguttzeit/AI-Code-Examination}; Analysis Repository:  \url{https://github.com/t-muras/AI-Code-Analysis}.

\section{Statistical Analysis} \label{app:statistical-analysis}
This section contains additional information that outlines our study's findings. Frequency tables are used to provide precise data on the quality and correctness of the produced Python and Java code. Furthermore, all results from hypothesis tests and statistical analysis are presented in detail.

\subsection{Frequency Tables for Correctness}
The tables below display the count of generations per AI model and in case of test generation per approach, categorized as \textit{Correct} or \textit{Incorrect}, based on the unit test outcomes for a particular generation. 
\paragraph{Source Code Generation}
\begin{figure}[H]
    \centering
    \vspace{-10px}
    \begin{minipage}[c]{.49\textwidth}
\begin{table}[H]
    \footnotesize     \raggedright 
    \begin{tabular}{lccc}
    \headrow
    \thead{} & \thead{Correct} & \thead{Incorrect} & \thead{Total generations}\\
    ChatGPT        & 536                 & 64                    & 600 \\ 
    GitHub Copilot & 453                 & 147                   & 600 \\
    \hline
    \end{tabular} \\
    \caption{Frequency table of Code Correctness results for Java}
\end{table}
    \end{minipage}
    \begin{minipage}[c]{.49\textwidth}
\begin{table}[H]
    \footnotesize     \raggedright
    \begin{tabular}{lccc}
    \headrow
    \thead{} & \thead{Correct} & \thead{Incorrect} & \thead{Total generations}\\
    ChatGPT        & 475                 & 125                    & 600 \\ 
    GitHub Copilot & 375                 & 225                   & 600 \\
    \hline
    \end{tabular} \\
    \caption{Frequency table of Code Correctness results for Python}
\end{table}
    \end{minipage}
\end{figure}

\paragraph{Test Case Generation}
\begin{table}[H]
    \footnotesize     \raggedright
    \begin{tabular}{llccc}
    \headrow
    \thead{} & \thead{Approach} & \thead{Correct} & \thead{Incorrect} & \thead{Total generations}\\
    \textbf{ChatGPT} & AIGenerated     & 50                 & 70                    & 120 \\
            & BookExampleCode & 44                 & 76                    & 120 \\
            & PromptOnly      & 41                 & 79                    & 120 \\
    \textbf{GitHub Copilot} & AIGenerated     & 64                 & 56                   & 120 \\
                   & BookExampleCode & 61                 & 59                    & 120 \\
                   & PromptOnly      & 54                 & 66                    & 120 \\
    \hline
    \end{tabular} \\
    \caption{Frequency table of Test Correctness results for Java}
\end{table}

\begin{table}[H]
    \footnotesize     \raggedright
    \begin{tabular}{llccc}
    \headrow
    \thead{} & \thead{Approach} & \thead{Correct} & \thead{Incorrect} & \thead{Total generations}\\
    \textbf{ChatGPT} & AIGenerated     & 40                 & 80                    & 120 \\
            & BookExampleCode & 30                 & 90                    & 120 \\
            & PromptOnly      & 33                 & 87                    & 120 \\
    \textbf{GitHub Copilot} & AIGenerated     & 39                 & 81                   & 120 \\
                   & BookExampleCode & 48                 & 72                    & 120 \\
                   & PromptOnly      & 54                 & 66                    & 120 \\
    \hline
    \end{tabular} \\
    \caption{Frequency table of Test Correctness results for Python}
\end{table}

\subsection{Frequency Tables for Code Quality}
The following tables show the number of lines of code generated per AI model, classified into \textit{lines with no errors} and \textit{lines with errors}. The number of lines is calculated across all twelve algorithms. Additionally, they indicate the quantity of quality rule violations for each AI model, according to the quality rules defined in section~\ref{sec:code-quality}.
\paragraph{Source Code Generation in Java}
\begin{table}[H]
    \footnotesize     \raggedright
    \begin{tabular}{lccc}
    \headrow
    \thead{} & \thead{Lines without errors} & \thead{Lines with errors} & \thead{Total number of Lines}\\
    ChatGPT        & 20~263                 & 394                    & 20~657 \\ 
    GitHub Copilot & 19~186                 & 366                   & 19~552 \\
    \hline
    \end{tabular} \\
    \caption{Frequency table of Code Quality results for Java}
\end{table}

\begin{table}[H]
    \footnotesize     \raggedright
    \begin{tabular}{lccccc}
    \headrow
    \thead{Number of quality rule violations} & \thead{0} & \thead{1} & \thead{2} & \thead{3} & \thead{4}\\
    ChatGPT        & 359 & 139 & 57 & 39 & 6 \\ 
    GitHub Copilot & 343 & 170 & 66 & 20 & 1 \\
    \hline
    \end{tabular} \\
    \caption{Frequency table of Quality rule violations for Java}
\end{table}

\paragraph{Source Code Generation in Python}
\begin{table}[H]
    \footnotesize     \raggedright
    \begin{tabular}{lccc}
    \headrow
    \thead{} & \thead{Lines without errors} & \thead{Lines with errors} & \thead{Total number of Lines}\\
    ChatGPT        & 10~700                 & 1~432                  & 12~132 \\ 
    GitHub Copilot & 10~768                 & 1~177                  & 11~945 \\
    \hline
    \end{tabular} \\
    \caption{Frequency table of Code Quality results for Python}
\end{table}

\begin{table}[H]
    \footnotesize     \raggedright
    \begin{tabular}{lcccccccccccc}
    \headrow
    \thead{Number of quality rule violations} & \thead{0} & \thead{1} & \thead{2} & \thead{3} & \thead{4} & \thead{5} & \thead{6} & \thead{7} & \thead{8} & \thead{9} & \thead{10} & \thead{11}\\
    ChatGPT        & 127 & 142 & 73 & 114 & 41 & 35 & 34 & 22 & 5 & 6 & 0 & 1\\ 
    GitHub Copilot & 158 & 148 & 61 & 94 & 89 & 36 & 11 & 2 & 0 & 1 &  & \\
    \hline
    \end{tabular} \\
    \caption{Frequency table of Quality rule violations for Python}
\end{table}

\subsection{Hypothesis Tests}
As previously mentioned, several hypothesis tests were performed to identify statistically significant differences in AI tools, languages and test case generation approaches. The following section displays the precise results of the tests alongside corresponding links to the AI-Therapy website, where they were conducted. The Mann-Whitney U test was employed to compare two groups, while the Kruskal Wallis test was used to compare the three test generation approaches. Both tests aim to evaluate group distributions by testing the null hypothesis, which implies no differences between them. The alternative hypothesis suggests that such differences do exist.
\subsubsection{Source Code Generation}
\paragraph{Code Correctness}
The following tables show the results of the hypothesis tests used to examine the \textbf{differences in the code correctness between ChatGPT and GitHub Copilot}. These are presented separately for Java and Python:
\begin{figure}[H]
    \centering
    \vspace{-10px}
    \begin{minipage}[c]{.49\textwidth}
        \begin{table}[H]
\footnotesize
            \centering
            \begin{tabular}{lr}
            \headrow
            \thead{\href{https://www.ai-therapy.com/psychology-statistics/results/20231111102137758}{Mann-Whitney U Test}} & \thead{Java} \\
            U             & 155100.000 \\ 
            z-score       & -6.291 \\
            p             & < 0.001 \\
            effect size r & -0.182 \\
            \hline
            \end{tabular} \\
            \captionsetup{justification=centering}
            \caption{Mann-Whitney U Test comparing code correctness results of the AI models for Java}
        \end{table}
    \end{minipage}
    \begin{minipage}[c]{.49\textwidth}
        \begin{table}[H]
\footnotesize
            \centering 
            \begin{tabular}{lr}
            \headrow
            \thead{\href{https://www.ai-therapy.com/psychology-statistics/results/20231111102320935}{Mann-Whitney U Test}} & \thead{Python} \\
            U             & 150000.000 \\ 
            z-score       & -6.348 \\
            p             & < 0.001 \\
            effect size r & -0.183 \\
            \hline
            \end{tabular} \\
            \captionsetup{justification=centering}
            \caption{Mann-Whitney U Test comparing code correctness results of the AI models for Python}
        \end{table}
    \end{minipage}
\end{figure}

The following tables present hypothesis test outcomes evaluating \textbf{differences between the programming languages Java and Python}, for both ChatGPT and GitHub Copilot individually:
\begin{figure}[H]
    \centering
    \vspace{-10px}
    \begin{minipage}[c]{.49\textwidth}
        \vspace{-11px}
        \begin{table}[H]
\footnotesize
            \centering
            \begin{tabular}{lr}
            \headrow
            \thead{\href{https://www.ai-therapy.com/psychology-statistics/results/20231111102532634}{Mann-Whitney U Test}} & \thead{ChatGPT} \\
            U             & 161700.000 \\ 
            z-score       & -4.832 \\
            p             & < 0.001 \\
            effect size r & -0.139 \\
            \hline
            \end{tabular} \\
            \captionsetup{justification=centering}
            \caption{Mann-Whitney U Test comparing code correctness results of Java and Python for ChatGPT}
        \end{table}
    \end{minipage}
    \begin{minipage}[c]{.49\textwidth}
        \begin{table}[H]
\footnotesize
            \centering 
            \begin{tabular}{lr}
            \headrow
            \thead{\href{https://www.ai-therapy.com/psychology-statistics/results/20230922143540378}{Mann-Whitney U Test}} & \thead{GitHub Copilot} \\
            U             & 156600.000 \\ 
            z-score       & -4.867 \\
            p             & < 0.001 \\
            effect size r & -0.140 \\
            \hline
            \end{tabular} \\
            \captionsetup{justification=centering}
            \caption{Mann-Whitney U Test comparing code correctness results of Java and Python for GitHub Copilot}
        \end{table}
    \end{minipage}
\end{figure}

\paragraph{Code Quality}
The following tables show the results of the hypothesis tests used to examine the \textbf{differences in the code quality between ChatGPT and GitHub Copilot}. These are presented separately for Java and Python:
\begin{figure}[h]
    \centering
    \vspace{-10px}
    \begin{minipage}[c]{.49\textwidth}
        \begin{table}[H]
\footnotesize
            \centering
            \begin{tabular}{lr}
            \headrow
            \thead{\href{https://www.ai-therapy.com/psychology-statistics/results/20230922130117040}{Mann-Whitney U Test}} & \thead{Java} \\
            U             & 178750.500 \\ 
            z-score       & -0.235 \\
            p             & 0.814 \\
            effect size r & -0.007 \\
            \hline
            \end{tabular} \\
            \captionsetup{justification=centering}
            \caption{Mann-Whitney U Test comparing Code Quality results of the AI models for Java}
        \end{table}
    \end{minipage}
    \begin{minipage}[c]{.49\textwidth}
        \begin{table}[H]
\footnotesize
            \centering 
            \begin{tabular}{lr}
            \headrow
            \thead{\href{https://www.ai-therapy.com/psychology-statistics/results/20231111102713209}{Mann-Whitney U Test}} & \thead{Python} \\
            U             & 164198.000 \\ 
            z-score       & -2.681 \\
            p             & 0.007 \\
            effect size r & -0.077 \\
            \hline
            \end{tabular} \\
            \captionsetup{justification=centering}
            \caption{Mann-Whitney U Test comparing Code Quality results of the AI models for Python}
        \end{table}
    \end{minipage}
\end{figure}

The following tables present hypothesis test outcomes evaluating \textbf{differences between the programming languages Java and Python}, for both ChatGPT and GitHub Copilot individually:
\begin{figure}[h]
    \centering
    \vspace{-10px}
    \begin{minipage}[c]{.49\textwidth}
        \begin{table}[H]
\footnotesize
            \centering
            \begin{tabular}{lr}
            \headrow
            \thead{\href{https://www.ai-therapy.com/psychology-statistics/results/20230922141134585}{Mann-Whitney U Test}} & \thead{ChatGPT} \\
            U             & 86152.000 \\ 
            z-score       & -16.325 \\
            p             & < 0.001 \\
            effect size r & -0.471 \\
            \hline
            \end{tabular} \\
            \captionsetup{justification=centering}
            \caption{Mann-Whitney U Test comparing Code Quality results of Java and Python for ChatGPT}
        \end{table}
    \end{minipage}
    \begin{minipage}[c]{.49\textwidth}
        \begin{table}[H]
\footnotesize
            \centering 
            \begin{tabular}{lr}
            \headrow
            \thead{\href{https://www.ai-therapy.com/psychology-statistics/results/20230922141326709}{Mann-Whitney U Test}} & \thead{GitHub Copilot} \\
            U             & 97531.500 \\ 
            z-score       & -14.433 \\
            p             & < 0.001 \\
            effect size r & -0.417 \\
            \hline
            \end{tabular} \\
            \captionsetup{justification=centering}
            \caption{Mann-Whitney U Test comparing Code Quality results of Java and Python for GitHub Copilot}
        \end{table}
    \end{minipage}
\end{figure}

\subsubsection{Test Case Generation}

\paragraph{Code Correctness}
The following tables show the hypothesis tests results evaluating the \textbf{differences between the three approaches for the programming language Java}. These are presented separately for ChatGPT and GitHub Copilot:
\begin{figure}[H]
    \centering
    \vspace{-10px}
    \begin{minipage}[c]{.49\textwidth}
        \begin{table}[H]
\footnotesize
            \centering
            \begin{tabular}{lr}
            \headrow
            \thead{\href{https://www.ai-therapy.com/psychology-statistics/results/20231111103410983}{Kruskal-Wallis Test}} & \thead{Java - ChatGPT} \\
            H(2)          & 1.489 \\ 
            p             & 0.475 \\
            \hline
            \end{tabular} \\
            \captionsetup{justification=centering}
            \caption{Kruskal-Wallis Test comparing code correctness results of the three approaches for Java using ChatGPT}
        \end{table}
    \end{minipage}
    \begin{minipage}[c]{.49\textwidth}
        \begin{table}[H]
\footnotesize
            \centering 
            \begin{tabular}{lr}
            \headrow
            \thead{\href{https://www.ai-therapy.com/psychology-statistics/results/20230926183159450}{Kruskal-Wallis Test}} & \thead{Java - GitHub Copilot} \\
            H(2)          & 1.751 \\ 
            p             & 0.417 \\
            \hline
            \end{tabular} \\
            \captionsetup{justification=centering}
            \caption{Kruskal-Wallis Test comparing code correctness results of the three approaches for Java using GitHub Copilot}
        \end{table}
    \end{minipage}
\end{figure}

The following tables show the hypothesis tests results evaluating the \textbf{differences between the three approaches for the programming language Python}. These are presented separately for ChatGPT and GitHub Copilot:
\begin{figure}[H]
    \centering
    \vspace{-10px}
    \begin{minipage}[c]{.49\textwidth}
        \begin{table}[H]
\footnotesize
            \centering
            \begin{tabular}{lr}
            \headrow
            \thead{\href{https://www.ai-therapy.com/psychology-statistics/results/20231111103705019}{Kruskal-Wallis Test}} & \thead{Python - ChatGPT} \\
            H(2)          & 2.143\\ 
            p             & 0.343 \\
            \hline
            \end{tabular} \\
            \captionsetup{justification=centering}
            \caption{Kruskal-Wallis Test comparing code correctness results of the three approaches for Python using ChatGPT}
        \end{table}
    \end{minipage}
    \begin{minipage}[c]{.49\textwidth}
        \begin{table}[H]
\footnotesize
            \centering 
            \begin{tabular}{lr}
            \headrow
            \thead{\href{https://www.ai-therapy.com/psychology-statistics/results/20230926203125779}{Kruskal-Wallis Test}} & \thead{Python - GitHub Copilot} \\
            H(2)          & 3.976 \\ 
            p             & 0.137 \\
            \hline
            \end{tabular} \\
            \captionsetup{justification=centering}
            \caption{Kruskal-Wallis Test comparing code correctness results of the three approaches for Python using GitHub Copilot}
        \end{table}
    \end{minipage}
\end{figure}

The presented tables display the outcomes of hypothesis testing that investigated the \textbf{variation in code correctness between ChatGPT and GitHub Copilot}, combining all three approaches. These are presented separately for Java and Python:
\begin{figure}[H]
    \centering
    \vspace{-10px}
    \begin{minipage}[c]{.49\textwidth}
        \begin{table}[H]
\footnotesize
            \centering
            \begin{tabular}{lr}
            \headrow
            \thead{\href{https://www.ai-therapy.com/psychology-statistics/results/20230926200044073}{Mann-Whitney U Test}} & \thead{Java} \\
            U             & 56880.000 \\ 
            z-score       & -3.304 \\
            p             & < 0.001 \\
            effect size r & -0.123 \\
            \hline
            \end{tabular} \\
            \captionsetup{justification=centering}
            \caption{Mann-Whitney U Test comparing code correctness results of the AI models for Java}
        \end{table}
    \end{minipage}
    \begin{minipage}[c]{.49\textwidth}
        \begin{table}[H]
\footnotesize
            \centering 
            \begin{tabular}{lr}
            \headrow
            \thead{\href{https://www.ai-therapy.com/psychology-statistics/results/20230926204556824}{Mann-Whitney U Test}} & \thead{Python} \\
            U             & 57960.000 \\ 
            z-score       & -2.990 \\
            p             & 0.003 \\
            effect size r & -0.111 \\
            \hline
            \end{tabular} \\
            \captionsetup{justification=centering}
            \caption{Mann-Whitney U Test comparing code correctness results of the AI models for Python}
        \end{table}
    \end{minipage}
\end{figure}

The following tables present hypothesis test outcomes evaluating \textbf{differences between the programming languages Java and Python}, combined for ChatGPT and GitHub Copilot:
\begin{table}[H]
\footnotesize
            \centering
            \begin{tabular}{lr}
            \headrow
            \thead{\href{https://www.ai-therapy.com/psychology-statistics/results/20230926210215349}{Mann-Whitney U Test}} & \thead{Correctness} \\
            U             & 234000.000 \\ 
            z-score       & -3.785 \\
            p             & < 0.001 \\
            effect size r & -0.100 \\
            \hline
            \end{tabular} \\
            \captionsetup{justification=centering}
            \caption{Mann-Whitney U Test comparing code correctness results of Java and Python for both AI models}
\end{table}

\paragraph{Coverage}
The following tables show the hypothesis tests results, evaluating the \textbf{differences in the coverage between the three approaches for the programming language Java}. These are presented separately for ChatGPT and GitHub Copilot:
\begin{figure}[H]
    \centering
    \vspace{-10px}
    \begin{minipage}[c]{.49\textwidth}
        \begin{table}[H]
\footnotesize
            \centering
            \begin{tabular}{lr}
            \headrow
            \thead{\href{https://www.ai-therapy.com/psychology-statistics/results/20231111104059593}{Kruskal-Wallis Test}} & \thead{Java - ChatGPT} \\
            H(2)          & 1.799 \\ 
            p             & 0.407 \\
            \hline
            \end{tabular} \\
            \captionsetup{justification=centering}
            \caption{Kruskal-Wallis Test comparing coverage results of the three approaches for Java using ChatGPT}
        \end{table}
    \end{minipage}
    \begin{minipage}[c]{.49\textwidth}
        \begin{table}[H]
\footnotesize
            \centering 
            \begin{tabular}{lr}
            \headrow
            \thead{\href{https://www.ai-therapy.com/psychology-statistics/results/20231111104309487}{Kruskal-Wallis Test}} & \thead{Java - GitHub Copilot} \\
            H(2)          & 6.769 \\ 
            p             & 0.034 \\
            \hline
            \end{tabular} \\
            \captionsetup{justification=centering}
            \caption{Kruskal-Wallis Test comparing coverage results of the three approaches for Java using GitHub Copilot}
        \end{table}
    \end{minipage}
\end{figure}

The following tables show the hypothesis tests results evaluating the \textbf{differences between the three approaches for the programming language Python}. These are presented separately for ChatGPT and GitHub Copilot:
\begin{figure}[H]
    \centering
    \vspace{-10px}
    \begin{minipage}[c]{.49\textwidth}
        \begin{table}[H]
\footnotesize
            \centering
            \begin{tabular}{lr}
            \headrow
            \thead{\href{https://www.ai-therapy.com/psychology-statistics/results/20231111104546431}{Kruskal-Wallis Test}} & \thead{Python - ChatGPT} \\
            H(2)          & 39.905\\ 
            p             & < 0.001 \\
            \hline
            \end{tabular} \\
            \captionsetup{justification=centering}
            \caption{Kruskal-Wallis Test comparing coverage results of the three approaches for Python using ChatGPT}
        \end{table}
    \end{minipage}
    \begin{minipage}[c]{.49\textwidth}
        \begin{table}[H]
\footnotesize
            \centering 
            \begin{tabular}{lr}
            \headrow
            \thead{\href{https://www.ai-therapy.com/psychology-statistics/results/20230926215447516}{Kruskal-Wallis Test}} & \thead{Python - GitHub Copilot} \\
            H(2)          & 23.861 \\ 
            p             & < 0.001 \\
            \hline
            \end{tabular} \\
            \captionsetup{justification=centering}
            \caption{Kruskal-Wallis Test comparing coverage results of the three approaches for Python using GitHub Copilot}
        \end{table}
    \end{minipage}
\end{figure}

The presented tables display the outcomes of hypothesis testing that investigated the \textbf{variation in coverage between ChatGPT and GitHub Copilot}, combining all three approaches. These are presented separately for Java and Python:
\begin{figure}[H]
    \centering
    \vspace{-10px}
    \begin{minipage}[c]{.49\textwidth}
        \begin{table}[H]
\footnotesize
            \centering
            \begin{tabular}{lr}
            \headrow
            \thead{\href{https://www.ai-therapy.com/psychology-statistics/results/20230926214810412}{Mann-Whitney U Test}} & \thead{Java} \\
            U             & 57456.000 \\ 
            z-score       & -2.635 \\
            p             & 0.008 \\
            effect size r & -0.098 \\
            \hline
            \end{tabular} \\
            \captionsetup{justification=centering}
            \caption{Mann-Whitney U Test comparing coverage results of the AI models for Java}
        \end{table}
    \end{minipage}
    \begin{minipage}[c]{.49\textwidth}
        \begin{table}[H]
\footnotesize
            \centering 
            \begin{tabular}{lr}
            \headrow
            \thead{\href{https://www.ai-therapy.com/psychology-statistics/results/20231111104802326}{Mann-Whitney U Test}} & \thead{Python} \\
            U             & 62699.000 \\ 
            z-score       & -0.857 \\
            p             & 0.391 \\
            effect size r & -0.032 \\
            \hline
            \end{tabular} \\
            \captionsetup{justification=centering}
            \caption{Mann-Whitney U Test comparing coverage results of the AI models for Python}
        \end{table}
    \end{minipage}
\end{figure}

The following tables present hypothesis test outcomes evaluating \textbf{differences between the programming languages Java and Python}, combined for ChatGPT and GitHub Copilot:
\begin{table}[H]
\footnotesize
            \centering
            \begin{tabular}{lr}
            \headrow
            \thead{\href{https://www.ai-therapy.com/psychology-statistics/results/20230926215858966}{Mann-Whitney U Test}} & \thead{Coverage} \\
            U             & 85834.500 \\ 
            z-score       & -22.297 \\
            p             & < 0.001 \\
            effect size r & -0.588 \\
            \hline
            \end{tabular} \\
            \captionsetup{justification=centering}
            \caption{Mann-Whitney U Test comparing coverage results of Java and Python for both AI models}
\end{table}

\paragraph{Modification Rate}
The presented tables display the outcomes of hypothesis testing that investigated the \textbf{differences in the Levenshtein Distance between ChatGPT and GitHub Copilot}, combining all three approaches. These are presented separately for Java and Python:
\begin{figure}[H]
    \centering
    \vspace{-10px}
    \begin{minipage}[c]{.49\textwidth}
        \begin{table}[H]
\footnotesize
            \centering
            \begin{tabular}{lr}
            \headrow
            \thead{\href{https://www.ai-therapy.com/psychology-statistics/results/20230926211700625}{Mann-Whitney U Test}} & \thead{Java} \\
            U             & 40589.500 \\ 
            z-score       & -11.212 \\
            p             & < 0.001 \\
            effect size r & -0.418 \\
            \hline
            \end{tabular} \\
            \captionsetup{justification=centering}
            \caption{Mann-Whitney U Test comparing Levenshtein Distance results of the AI models for Java}
        \end{table}
    \end{minipage}
    \begin{minipage}[c]{.49\textwidth}
        \begin{table}[H]
\footnotesize
            \centering 
            \begin{tabular}{lr}
            \headrow
            \thead{\href{https://www.ai-therapy.com/psychology-statistics/results/20230926212806271}{Mann-Whitney U Test}} & \thead{Python} \\
            U             & 45498.500 \\ 
            z-score       & -8.974\\
            p             & < 0.001 \\
            effect size r & -0.334 \\
            \hline
            \end{tabular} \\
            \captionsetup{justification=centering}
            \caption{Mann-Whitney U Test comparing Levenshtein Distance results of the AI models for Python}
        \end{table}
    \end{minipage}
\end{figure}

The following tables present hypothesis test outcomes evaluating \textbf{differences between the programming languages Java and Python}, combined for ChatGPT and GitHub Copilot:
\begin{table}[H]
\footnotesize
            \centering
            \begin{tabular}{lr}
            \headrow
            \thead{\href{https://www.ai-therapy.com/psychology-statistics/results/20231111105046468}{Mann-Whitney U Test}} & \thead{Levenshtein Distance} \\
            U             & 259117.500 \\ 
            z-score       & -0.014 \\
            p             & 0.989 \\
            effect size r & -0.000 \\
            \hline
            \end{tabular} \\
            \captionsetup{justification=centering}
            \caption{Mann-Whitney U Test comparing Levenshtein Distance results of Java and Python for both AI models}
\end{table}

\subsubsection{Comparison with baseline study}
The following tables show the results of the hypothesis tests used to examine the \textbf{differences in the code correctness between our results and the ones of the baseline study} using the same six Java algorithms. These are presented separately for ChatGPT and GitHub Copilot:
\begin{figure}[H]
    \centering
    \vspace{-10px}
    \begin{minipage}[c]{.49\textwidth}
        \begin{table}[H]
\footnotesize
            \centering
            \begin{tabular}{lr}
            \headrow
            \thead{\href{https://www.ai-therapy.com/psychology-statistics/results/20230916113200481}{Mann-Whitney U Test}} & \thead{ChatGPT} \\
            U             & 39600.000 \\ 
            z-score       & -5.887 \\
            p             & < 0.001 \\
            effect size r & -0.240 \\
            \hline
            \end{tabular} \\
            \captionsetup{justification=centering}
            \caption{Mann-Whitney U Test comparing code correctness results of the baseline study and current outcomes for ChatGPT}
        \end{table}
    \end{minipage}
    \begin{minipage}[c]{.49\textwidth}
        \begin{table}[H]
\footnotesize
            \centering 
            \begin{tabular}{lr}
            \headrow
            \thead{\href{https://www.ai-therapy.com/psychology-statistics/results/20231111215121702}{Mann-Whitney U Test}} & \thead{GitHub Copilot} \\
            U             & 44700.000 \\ 
            z-score       & -0.257 \\
            p             & 0.797 \\
            effect size r & -0.011 \\
            \hline
            \end{tabular} \\
            \captionsetup{justification=centering}
            \caption{Mann-Whitney U Test comparing code correctness results of the baseline study and current outcomes for GitHub Copilot}
        \end{table}
    \end{minipage}
\end{figure}

The following tables present hypothesis test outcomes evaluating \textbf{differences between the results of our study and those of the baseline research}, for both ChatGPT and GitHub Copilot individually:
\begin{figure}[h]
    \centering
    \vspace{-10px}
    \begin{minipage}[c]{.49\textwidth}
        \begin{table}[H]
\footnotesize
            \centering
            \begin{tabular}{lr}
            \headrow
            \thead{\href{https://www.ai-therapy.com/psychology-statistics/results/20230922122227067}{Mann-Whitney U Test}} & \thead{ChatGPT} \\
            U             & 41966.000 \\ 
            z-score       & -1.947 \\
            p             & 0.052 \\
            effect size r & -0.079 \\
            \hline
            \end{tabular} \\
            \captionsetup{justification=centering}
            \caption{Mann-Whitney U Test comparing code quality results of the baseline study and current outcomes for ChatGPT}
        \end{table}
    \end{minipage}
    \begin{minipage}[c]{.49\textwidth}
        \begin{table}[H]
\footnotesize
            \centering 
            \begin{tabular}{lr}
            \headrow
            \thead{\href{https://www.ai-therapy.com/psychology-statistics/results/20230922122522604}{Mann-Whitney U Test}} & \thead{GitHub Copilot} \\
            U             & 41025.000 \\ 
            z-score       & -2.243 \\
            p             & 0.025 \\
            effect size r & -0.092\\
            \hline
            \end{tabular} \\
            \captionsetup{justification=centering}
            \caption{Mann-Whitney U Test comparing code quality results of the baseline study and current outcomes for GitHub Copilot}
        \end{table}
    \end{minipage}
\end{figure}

\subsection{Analysis with AI-Therapy}
The links to the AI-Therapy website, where the statistical analysis was conducted, are provided below. Each metric for both ChatGPT and GitHub Copilot was subjected to mean, mode, median, dispersion, and normality tests separately. 
\subsubsection{Source Code Generation}
\paragraph{Code Correctness}
\begin{table}[H]
\footnotesize
    \centering
    \begin{tabular}{p{0.13\textwidth}p{0.38\textwidth}p{0.38\textwidth}}
    \headrow
    \thead{} & \thead{ChatGPT} & \thead{Copilot}\\
    Mean, Mode \& Median &  \url{https://www.ai-therapy.com/psychology-statistics/results/20230922144727108}                &   \url{https://www.ai-therapy.com/psychology-statistics/results/20230922144755056}  \\ 
    Dispersion           &  \url{https://www.ai-therapy.com/psychology-statistics/results/20230922144704553}                &  \url{https://www.ai-therapy.com/psychology-statistics/results/20230922144813189}  \\
    Normality test       &  \url{https://www.ai-therapy.com/psychology-statistics/results/20230922144629088}                &   \url{https://www.ai-therapy.com/psychology-statistics/results/20230922144850536}    \\
    \hline
    \end{tabular} \\
    \captionsetup{justification=centering}
    \caption{Statistical Analysis for Code Correctness of Java code}
\end{table}

\begin{table}[H]
\footnotesize
    \centering
    \begin{tabular}{p{0.13\textwidth}p{0.38\textwidth}p{0.38\textwidth}}
    \headrow
    \thead{} & \thead{ChatGPT} & \thead{Copilot}\\
    Mean, Mode \& Median &  \url{https://www.ai-therapy.com/psychology-statistics/results/20230922144434667} &    \url{https://www.ai-therapy.com/psychology-statistics/results/20230922144151681} \\ 
    Dispersion           & \url{https://www.ai-therapy.com/psychology-statistics/results/20230922144411895} & \url{https://www.ai-therapy.com/psychology-statistics/results/20230922144219794} \\
    Normality test       & \url{https://www.ai-therapy.com/psychology-statistics/results/20230922144347165} &  \url{https://www.ai-therapy.com/psychology-statistics/results/20230922144256711}  \\
    \hline
    \end{tabular} \\
    \captionsetup{justification=centering}
    \caption{Statistical Analysis for Code Correctness of Python code}
\end{table}

\paragraph{Code Quality}
\begin{table}[H]
\footnotesize
    \centering
    \begin{tabular}{p{0.13\textwidth}p{0.38\textwidth}p{0.38\textwidth}}
    \headrow
    \thead{} & \thead{ChatGPT} & \thead{Copilot}\\
    Mean, Mode \& Median &  \url{https://www.ai-therapy.com/psychology-statistics/results/20230922125457835} & \url{https://www.ai-therapy.com/psychology-statistics/results/20230922125911300}  \\ 
    Dispersion           & \url{https://www.ai-therapy.com/psychology-statistics/results/20230922125614192} & \url{https://www.ai-therapy.com/psychology-statistics/results/20230922125831227} \\
    Normality test       &  \url{https://www.ai-therapy.com/psychology-statistics/results/20230922125721837} & \url{https://www.ai-therapy.com/psychology-statistics/results/20230922125806193}   \\
    \hline
    \end{tabular} \\
    \captionsetup{justification=centering}
    \caption{Statistical Analysis for Code Quality of Java code}
\end{table}

\begin{table}[H]
\footnotesize
    \centering
    \begin{tabular}{p{0.13\textwidth}p{0.38\textwidth}p{0.38\textwidth}}
    \headrow
    \thead{} & \thead{ChatGPT} & \thead{Copilot}\\
    Mean, Mode \& Median &  \url{https://www.ai-therapy.com/psychology-statistics/results/20230922135305224} & \url{https://www.ai-therapy.com/psychology-statistics/results/20230922135749242}  \\ 
    Dispersion           & \url{https://www.ai-therapy.com/psychology-statistics/results/20230922135429913} & \url{https://www.ai-therapy.com/psychology-statistics/results/20230922135657746} \\
    Normality test       &  \url{https://www.ai-therapy.com/psychology-statistics/results/20230922135540031} & \url{https://www.ai-therapy.com/psychology-statistics/results/20230922135629920}   \\
    \hline
    \end{tabular} \\
    \captionsetup{justification=centering}
    \caption{Statistical Analysis for Code Quality of Python code}
\end{table}

\subsubsection{Test Case Generation}
\paragraph{Code Correctness}
\begin{table}[H]
\footnotesize
    \centering
    \begin{tabular}{p{0.13\textwidth}p{0.38\textwidth}p{0.38\textwidth}}
    \headrow
    \thead{} & \thead{ChatGPT} & \thead{Copilot}\\
    Mean, Mode \& Median &  \url{https://www.ai-therapy.com/psychology-statistics/results/20231014174004808}                &   \url{https://www.ai-therapy.com/psychology-statistics/results/20231014174340691}  \\ 
    Dispersion           &  \url{https://www.ai-therapy.com/psychology-statistics/results/20231014174043166}                &  \url{https://www.ai-therapy.com/psychology-statistics/results/20231014174317297}  \\
    Normality test       &  \url{https://www.ai-therapy.com/psychology-statistics/results/20231014174211684}                &   \url{https://www.ai-therapy.com/psychology-statistics/results/20231014174258521}    \\
    \hline
    \end{tabular} \\
    \captionsetup{justification=centering}
    \caption{Statistical Analysis for Code Correctness of Java code}
\end{table}

\begin{table}[H]
\footnotesize
    \centering
    \begin{tabular}{p{0.13\textwidth}p{0.38\textwidth}p{0.38\textwidth}}
    \headrow
    \thead{} & \thead{ChatGPT} & \thead{Copilot}\\
    Mean, Mode \& Median &  \url{https://www.ai-therapy.com/psychology-statistics/results/20231014174635772}               &    \url{https://www.ai-therapy.com/psychology-statistics/results/20231014174845382} \\ 
    Dispersion           & \url{https://www.ai-therapy.com/psychology-statistics/results/20231014174659244} & \url{https://www.ai-therapy.com/psychology-statistics/results/20231014174823068} \\
    Normality test       & \url{https://www.ai-therapy.com/psychology-statistics/results/20231014174731212} &  \url{https://www.ai-therapy.com/psychology-statistics/results/20231014174800028}  \\
    \hline
    \end{tabular} \\
    \captionsetup{justification=centering}
    \caption{Statistical Analysis for Code Correctness of Python code}
\end{table}

\paragraph{Coverage}
\begin{table}[H]
\footnotesize
    \centering
    \begin{tabular}{p{0.13\textwidth}p{0.38\textwidth}p{0.38\textwidth}}
    \headrow
    \thead{} & \thead{ChatGPT} & \thead{Copilot}\\
    Mean, Mode \& Median &  \url{https://www.ai-therapy.com/psychology-statistics/results/20231014180218501}                &   \url{https://www.ai-therapy.com/psychology-statistics/results/20231014180338252}  \\ 
    Dispersion           &  \url{https://www.ai-therapy.com/psychology-statistics/results/20231014180239265}                &  \url{https://www.ai-therapy.com/psychology-statistics/results/20231014180356509}  \\
    Normality test       &  \url{https://www.ai-therapy.com/psychology-statistics/results/20231014180306590}                &   \url{https://www.ai-therapy.com/psychology-statistics/results/20231014180423421}    \\
    \hline
    \end{tabular} \\
    \captionsetup{justification=centering}
    \caption{Statistical Analysis for the Coverage of Java code}
\end{table}

\begin{table}[H]
\footnotesize
    \centering
    \begin{tabular}{p{0.13\textwidth}p{0.38\textwidth}p{0.38\textwidth}}
    \headrow
    \thead{} & \thead{ChatGPT} & \thead{Copilot}\\
    Mean, Mode \& Median &  \url{https://www.ai-therapy.com/psychology-statistics/results/20231014180928851}                &   \url{https://www.ai-therapy.com/psychology-statistics/results/20231014181408346}  \\ 
    Dispersion           &  \url{https://www.ai-therapy.com/psychology-statistics/results/20231014180952019}                &  \url{https://www.ai-therapy.com/psychology-statistics/results/20231014181309179}  \\
    Normality test       &  \url{https://www.ai-therapy.com/psychology-statistics/results/20231014181211619}                &   \url{https://www.ai-therapy.com/psychology-statistics/results/20231014181245043}    \\
    \hline
    \end{tabular} \\
    \captionsetup{justification=centering}
    \caption{Statistical Analysis for the Coverage of Python code}
\end{table}

\paragraph{Levenshtein Distance}
\begin{table}[H]
\footnotesize
    \centering
    \begin{tabular}{p{0.13\textwidth}p{0.38\textwidth}p{0.38\textwidth}}
    \headrow
    \thead{} & \thead{ChatGPT} & \thead{Copilot}\\
    Mean, Mode \& Median &  \url{https://www.ai-therapy.com/psychology-statistics/results/20231014175738530}                &   \url{https://www.ai-therapy.com/psychology-statistics/results/20231014180037960}  \\ 
    Dispersion           &  \url{https://www.ai-therapy.com/psychology-statistics/results/20231014175816836}                &  \url{https://www.ai-therapy.com/psychology-statistics/results/20231014180008955}  \\
    Normality test       &  \url{https://www.ai-therapy.com/psychology-statistics/results/20231014175906696}                &   \url{https://www.ai-therapy.com/psychology-statistics/results/20231014175942039}    \\
    \hline
    \end{tabular} \\
    \captionsetup{justification=centering}
    \caption{Statistical Analysis for the Levenshtein Distance of Java code}
\end{table}

\begin{table}[H]
\footnotesize
    \centering
    \begin{tabular}{p{0.13\textwidth}p{0.38\textwidth}p{0.38\textwidth}}
    \headrow
    \thead{} & \thead{ChatGPT} & \thead{Copilot}\\
    Mean, Mode \& Median &  \url{https://www.ai-therapy.com/psychology-statistics/results/20231014175114881}                &   \url{https://www.ai-therapy.com/psychology-statistics/results/20231014175536934}  \\ 
    Dispersion           &  \url{https://www.ai-therapy.com/psychology-statistics/results/20231014175144498}                &  \url{https://www.ai-therapy.com/psychology-statistics/results/20231014175513214}  \\
    Normality test       &  \url{https://www.ai-therapy.com/psychology-statistics/results/20231014175422664}                &   \url{https://www.ai-therapy.com/psychology-statistics/results/20231014175454264}    \\
    \hline
    \end{tabular} \\
    \captionsetup{justification=centering}
    \caption{Statistical Analysis for the Levenshtein Distance of Python code}
\end{table}

\footnotesize
\renewcommand{\baselinestretch}{0.75}

\section{Generation Prompts}\label{app:prompts}
\subsection{Source Code Generation}
\subsubsection{Java}
\paragraph{Bellman Ford}
\begin{verbatim}
// Implement a non-static BellmanFord class with a static method bellmanFord(List<Edge> edges, 
int source, int n). 
// The method should return a Map<Integer, Integer> which contains the shortest path to every 
other node in the graph. 
// If there is a negative cycle, return null. 
// If there is no path to another node, skip this node in the output. 
// Implement a public local static class Edge with public int parameters source, dest, weight. 
\end{verbatim}

\paragraph{Binary Search}
\begin{verbatim}
// Implement a non-static class named BinarySearch. 
// Implement the public binarySearch(int, int[]) method. The method should return a boolean.
\end{verbatim}

\paragraph{Binary To Decimal}
\begin{verbatim}
// Implement a non-static BinaryToDecimal class. 
// Implement the public convertToDecimal(String binary) method. The method should return an 
int with the decimal.
\end{verbatim}

\paragraph{Breadth First Search}
\begin{verbatim}
// Implement a non-static breadth first search class named Graph(int vertices). 
// Implement the public bfs(int sourceVertex) method. The method should return an integer 
array of parent nodes for each vertex in the graph. The array of parent node values are all
initialized to -1 in the bfs(int sourceVertex) method.
// Implement the public addEdge(int, int). The method should add an edge between two vertices.
\end{verbatim}

\paragraph{Depth First Search}
\begin{verbatim}
// Implement a non-static public class DepthFirstSearch. 
// Implement the global variable public ArrayList<Integer> visitedNodes. 
// Implement the global variable public List<List<Integer>>, which represents an adjacency list. 
// Implement a constructor public DepthFirstSearch(List<Edge> edges, int n). 
// Implement an inner class public static class Edge with global int variables source, dest. 
// Implement the public void dfs(int startNode, boolean[] discovered) method. The method 
should implement a depth first search algorithm by pre-order on an undirected graph and 
write the visited nodes into the global ArrayList<Integer> visitedNodes.
\end{verbatim}

\paragraph{Dijkstra}
\begin{verbatim}
// Implement a non-static Dijkstra class with a constructor Dijkstra(List<Edge> edges, int n). 
// Implement a public non-static method findShortestPaths(int source). 
// This method should return a Map<Integer, Integer> which contains the shortest path to
every other node in the graph. 
// If there is no path to another node, skip this node in the output. 
// Implement a public local static class Edge with public int parameters source, dest, weight. 
// Implement a public local static class Node with public int parameters vertex, weight. 
\end{verbatim}

\paragraph{Egyptian Fractions}
\begin{verbatim}
// Implement a non-static class named EgyptianFractions. 
// Implement the public build(Long numerator, Long denominator) method with a List<Long> return value. 
\end{verbatim}

\paragraph{Floyd Warshall}
\begin{verbatim}
// Implement a non-static FloydWarshall class with the constructor FloydWarshall(int  nodes). 
// Implement a non-static public method addEdge(int source, int dest, int weight), which returns nothing. 
// Implement a non-static public method path(int source, int dest), which returns an Integer List. 
// Implement a non-static public method run(), which runs the Floyd Warshall algorithm.
\end{verbatim}

\paragraph{Knapsack}
\begin{verbatim}
// Implement a non-static 0-1 knapsack class named Knapsack. 
// Implement the public int bottomUp(int capacity, int[] weights, int[] values) method. The 
method should return an int with the maximum value.
\end{verbatim}

\paragraph{Kruskal}
\begin{verbatim}
// Implement a non-static Kruskal class with a static method runKruskalAlgorithm(List edges,
int n). The method should return a List 
// Implement a public local static class Edge with public int parameters src, dest, weight. 
// Implement a public local static class DisjointSet which manages the graph.
\end{verbatim}

\paragraph{Merge Sort}
\begin{verbatim}
// Implement a non-static merge sort algorithm class named MergeSort with a public void mergeSort(int[]) method.
\end{verbatim}

\paragraph{Quick Sort}
\begin{verbatim}
// Implement a non-static quick sort algorithm class named QuickSort with a public void sort(int[]) method.
\end{verbatim}

\subsubsection{Python}
\paragraph{Bellman Ford}
\begin{verbatim}
# Implement a Bellman Ford algorithm class named BellmanFord in python.
# The constructor should take an integer "vertices".
# Implement a public add_edge() method which takes three integers "source", "destination" 
and "weight" and returns nothing.
# Implement a public method bellman_ford() which takes an integer "source".
# The method should return a dictionary, containing the shortest path to every other node in 
the graph. If there is a negative cycle, return None.
# If there is no path to another node, skip this node in the output.
\end{verbatim}

\paragraph{Binary Search}
\begin{verbatim}
# Implement a non-static class named BinarySearch in python.
# Implement the public binary_search() method.
# The method should use an integer, an integer list as parameters.
# The method should return a boolean.
\end{verbatim}

\paragraph{Binary To Decimal}
\begin{verbatim}
# Implement a non-static BinaryToDecimal class in python.
# Implement the public convert_to_decimal() method.
# The method should take a string "binary". The method should return an int with the decimal.
\end{verbatim}

\paragraph{Breadth First Search}
\begin{verbatim}
# Implement a non-static breadth first search class named Graph in python.
# The class should take an integer "vertices".
# Implement the public bfs() method. The method should take an integer "vertex".
# The method should return an integer list of visited nodes in the graph.
# Implement the public add_edge() method. The method should take two integers.
# The method should add an edge between two vertices.
\end{verbatim}

\paragraph{Depth First Search}
\begin{verbatim}
# Implement a depth first search algorithm class named DepthFirstSearch in python.
# Implement a public add_edge() method which takes two integers "source", "destination" and returns nothing.
# Implement a public dfs() method which takes an integer "start_node" and returns a list of visited nodes.
\end{verbatim}

\paragraph{Dijkstra}
\begin{verbatim}
# Implement a class named Dijkstra which implements the Dijkstra algorithm in python.
# The constructor should take an integer "vertices".
# Implement a method add_edge(), which takes three integers "source", "destination", "weight"
and returns nothing.
# Implement a method find_shortest_paths() which takes an integer "source" and returns a 
dictionary, containing the shortest path to every other node in the graph.
# If there is no path to another node, skip this node in the output.
\end{verbatim}

\paragraph{Egyptian Fractions}
\begin{verbatim}
# Implement an egyptian fraction algorithm class named EgyptianFractions in python.
# Implement the public build() method which takes two long "numerator", "denominator" and returns a list. 
\end{verbatim}

\paragraph{Floyd Warshall}
\begin{verbatim}
# Implement a FloydWarshall algorithm class named FloydWarshall in python.
# The constructor should take an integer "nodes".
# Implement a public method add_edge() which takes three integers "source", "destination", 
"weight" as parameters and returns nothing.
# Implement a public method path() which takes two integers "source" and "destination" and
returns an integer list. If there is no path from source to destination, return an empty list.
# Implement a public method run(), which runs the FloydWarshall algorithm.
\end{verbatim}

\paragraph{Knapsack}
\begin{verbatim}
# Implement a non-static 0-1 knapsack class named Knapsack in python.
# Implement the public bottom_up() method. The method should take an integer "capacity",
# an integer list "weights" and an integer list "values".
# The method should return an int with the maximum value.
\end{verbatim}

\paragraph{Kruskal}
\begin{verbatim}
# Implement a Kruskal algorithm class named Kruskal in python.
# The constructor should take an integer "vertices".
# Implement a public method add_edge() which takes three integers "source", "destination",
"weight" as parameters and returns nothing.
# Implement a method run_kruskal_algorithm() which creates the minimal spanning tree and
returns a list of edges.
\end{verbatim}

\paragraph{Merge Sort}
\begin{verbatim}
# Implement a non-static merge sort algorithm class named MergeSort with a public
# merge_sort() method in python. The method should take an integer list and
# should return the sorted list.
\end{verbatim}

\paragraph{Quick Sort}
\begin{verbatim}
# Implement a non-static quick sort algorithm class named QuickSort with a public sort()
# method in python. The method should take an integer list and should return nothing.
\end{verbatim}

\subsection{Test Case Generation}
\subsubsection{Java - Prompt Only}
\paragraph{Bellman Ford}
\begin{verbatim}
// Implement a java test class called BellmanFordTest using JUnit version 4.
// The class BellmanFord that is to be tested contains the following:
// A local public static class Edge with public int parameters source, dest, weight.
// A public static Map<Integer, Integer> bellmanFord(List<Edge> edges, int source, int n)
method. The returned Map contains the shortest path to every other node in the graph.
// If there is a negative cycle, the method returns null.
// If there is no path to another node or if the source and destination node are the same,
this node is skipped in the output.
// Implement at least 3 different test cases.
// Build three graphs with the following edges and weights:
// Graph 1:
// List<BellmanFord.Edge> edges = Arrays.asList(
// new BellmanFord.Edge(0, 0, 1),
// new BellmanFord.Edge(1, 2, 3),
// new BellmanFord.Edge(1, 3, 2),
// new BellmanFord.Edge(1, 4, 2),
// new BellmanFord.Edge(3, 2, 5),
// new BellmanFord.Edge(4, 3, 3)
// );
// Graph 2:
// List<BellmanFord.Edge> edges = Arrays.asList(
// new BellmanFord.Edge(0, 1, -1),
// new BellmanFord.Edge(0, 2, 4),
// new BellmanFord.Edge(1, 2, 3),
// new BellmanFord.Edge(1, 4, -2),
// new BellmanFord.Edge(3, 2, 5),
// new BellmanFord.Edge(3, 1, 1),
// new BellmanFord.Edge(4, 3, -3)
// );
// Graph 3:
// List<BellmanFord.Edge> edges = Arrays.asList(
// new BellmanFord.Edge(0, 1, -1),
// new BellmanFord.Edge(0, 2, 4),
// new BellmanFord.Edge(1, 2, 3),
// new BellmanFord.Edge(1, 3, 2),
// new BellmanFord.Edge(1, 4, 2),
// new BellmanFord.Edge(3, 2, 5),
// new BellmanFord.Edge(3, 1, 1),
// new BellmanFord.Edge(4, 3, -3)
// );
\end{verbatim}

\paragraph{Binary Search}
\begin{verbatim}
// Implement a java test class called BinarySearchTest using JUnit version 4.
// The class BinarySearch that is to be tested contains the following:
// A public boolean binarySearch(int x, int[] sortedNumbers) method.
// Implement at least 3 different test cases.
\end{verbatim}

\paragraph{Binary To Decimal}
\begin{verbatim}
// Implement a java test class called BinaryToDecimalTest using JUnit version 4.
// The class BinaryToDecimal that is to be tested contains the following:
// A public int convertToDecimal(String binary) method. The returned int contains the decimal.
// Implement at least 3 different test cases.
\end{verbatim}

\paragraph{Breadth First Search}
\begin{verbatim}
// Implement a java test class called GraphTest using JUnit version 4.
// The class Graph that is to be tested implements the breadth first search algorithm 
and contains the following:
// A constructor Graph(int vertices).
// A public int[] bfs(int sourceVertex) method. The returned integer array contains the
parent nodes for each vertex in the graph. The values of the parent node array are all 
initialized to -1 in the bfs(int sourceVertex) method.
// A public void addEdge(int, int) method, which adds an edge between two vertices.
// Implement at least 3 different test cases.
// Build a graph with the following edges:
// breadthFirstSearch.addEdge(0, 1);
// breadthFirstSearch.addEdge(0, 2);
// breadthFirstSearch.addEdge(1, 3);
// breadthFirstSearch.addEdge(2, 3);
// breadthFirstSearch.addEdge(3, 4);
// breadthFirstSearch.addEdge(3, 5);
\end{verbatim}

\paragraph{Depth First Search}
\begin{verbatim}
// Implement a java test class called DepthFirstSearchTest using JUnit version 4.
// The class DepthFirstSearch that is to be tested contains the following:
// A global variable public ArrayList<Integer> visitedNodes.
// A global variable public List<List<Integer>>, which represents an adjacency list.
// A constructor public DepthFirstSearch(List<Edge> edges, int n).
// An inner public static class Edge with global int variables source, dest.
// A public void dfs(int startNode, boolean[] discovered) method. The method implements
a depth first search algorithm by pre-order on an undirected graph and writes the visited 
nodes into the global ArrayList<Integer> visitedNodes.
// Consider all possible permutations.
// Implement at least 3 different test cases.
// Build a graph with the following edges:
// List<DepthFirstSearch.Edge> edges = Arrays.asList(
// new DepthFirstSearch.Edge(0, 1),
// new DepthFirstSearch.Edge(0, 3),
// new DepthFirstSearch.Edge(1, 4),
// new DepthFirstSearch.Edge(2, 4),
// new DepthFirstSearch.Edge(2, 5),
// new DepthFirstSearch.Edge(3, 1),
// new DepthFirstSearch.Edge(4, 3),
// new DepthFirstSearch.Edge(5, 5),
// new DepthFirstSearch.Edge(6, 6)
// );
\end{verbatim}

\paragraph{Dijkstra}
\begin{verbatim}
// Implement a java test class called DijkstraTest using JUnit version 4.
// The class Dijkstra that is to be tested contains the following:
// A constructor Dijkstra(List<Edge> edges, int n).
// A local public static class Edge with public int parameters source, dest, weight.
// A local public class Node with public int parameters vertex, weight.
// A public boolean onlyPositiveEdgeCosts() method.
// A public Map<Integer, Integer> findShortestPaths(int source) method. The returned Map
contains the shortest path to every other node in the graph.
// If there is no path to another node or if the source and destination node are the same,
the node is skipped in the output.
// Implement at least 3 different test cases.
// Build three graphs with the following edges and weights:
// Graph 1:
// List<Dijkstra.Edge> edges = Arrays.asList(
// new Dijkstra.Edge(0, 1, 1),
// new Dijkstra.Edge(0, 2, 4),
// new Dijkstra.Edge(1, 2, 3),
// new Dijkstra.Edge(1, 3, 2),
// new Dijkstra.Edge(1, 4, 2),
// new Dijkstra.Edge(3, 2, 5),
// new Dijkstra.Edge(4, 3, 3)
// );
// Graph 2:
// List<Dijkstra.Edge> edges = Arrays.asList(
// new Dijkstra.Edge(0, 1, 1),
// new Dijkstra.Edge(0, 2, 4),
// new Dijkstra.Edge(1, 2, 3),
// new Dijkstra.Edge(1, 3, 2),
// new Dijkstra.Edge(1, 4, 2),
// new Dijkstra.Edge(3, 2, 5),
// new Dijkstra.Edge(3, 1, 1),
// new Dijkstra.Edge(4, 3, 3)
// );
// Graph 3:
// List<Dijkstra.Edge> edges = Arrays.asList(
// new Dijkstra.Edge(0, 0, 1),
// new Dijkstra.Edge(1, 2, 3),
// new Dijkstra.Edge(1, 3, 2),
// new Dijkstra.Edge(1, 4, 2),
// new Dijkstra.Edge(3, 2, 5),
// new Dijkstra.Edge(4, 3, 3)
// );
\end{verbatim}

\paragraph{Egyptian Fractions}
\begin{verbatim}
// Implement a java test class called EgyptianFractionsTest using JUnit version 4.
// The class EgyptianFractions that is to be tested contains the following:
// A public List<Long> build(Long numerator, Long denominator) method.
// Implement at least 3 different test cases.
\end{verbatim}

\paragraph{Floyd Warshall}
\begin{verbatim}
// Implement a java test class called FloydWarshallTest using JUnit version 4.
// The class FloydWarshall that is to be tested contains the following:
// A constructor FloydWarshall(int nodes).
// A non-static public void addEdge(int source, int dest, int weight) method.
// A non-static public List<Integer> path(int source, int dest) method.
// A non-static public void run() method, which runs the floyd warshall algorithm.
// If there is no path from source to destination or if the source and destination node 
are the same, the node is skipped in the output.
// Implement at least 3 different test cases.
// Build two graphs with the following edges and weights:
// Graph 1:
// floydWarshall.addEdge(0, 1, 10);
// floydWarshall.addEdge(0, 3, 5);
// floydWarshall.addEdge(1, 3, 2);
// floydWarshall.addEdge(1, 2, 1);
// floydWarshall.addEdge(2, 4, 4);
// floydWarshall.addEdge(3, 1, 3);
// floydWarshall.addEdge(3, 2, 9);
// floydWarshall.addEdge(3, 4, 2);
// floydWarshall.addEdge(4, 2, 6);
// Graph 2:
// floydWarshall.addEdge(0, 2, 5);
// floydWarshall.addEdge(1, 3, 5);
// floydWarshall.addEdge(2, 3, 4);
// floydWarshall.addEdge(3, 2, 4);
// floydWarshall.addEdge(4, 3, 4);
// floydWarshall.addEdge(2, 4, -2);
\end{verbatim}

\paragraph{Knapsack}
\begin{verbatim}
// Implement a java test class called KnapsackTest using JUnit version 4.
// The class Knapsack that is to be tested contains the following:
// A public int bottomUp(int capacity, int[] weights, int[] values) method. The returned 
int contains the maximum value.
// Implement at least 3 different test cases.
\end{verbatim}

\paragraph{Kruskal}
\begin{verbatim}
// Implement a java test class called KruskalTest using JUnit version 4.
// The class Kruskal that is to be tested contains the following:
// A public static List<Edge> runKruskalAlgorithm(List<Edge> edges, int n) method.
// A local public static class Edge with public int parameters src, dest, weight.
// A local public static class DisjointSet, which manages the graph.
// Implement at least 3 different test cases.
// Build two graphs with the following edges and weights:
// Graph 1:
// List<Kruskal.Edge> edges = Arrays.asList(
// new Kruskal.Edge(0, 1, 7),
// new Kruskal.Edge(1, 2, 8),
// new Kruskal.Edge(0, 3, 5),
// new Kruskal.Edge(1, 3, 9),
// new Kruskal.Edge(1, 4, 7),
// new Kruskal.Edge(2, 4, 5),
// new Kruskal.Edge(3, 4, 15),
// new Kruskal.Edge(3, 5, 6),
// new Kruskal.Edge(4, 5, 8),
// new Kruskal.Edge(4, 6, 9),
// new Kruskal.Edge(5, 6, 11)
// );
// Graph 2:
// List<Kruskal.Edge> edges = Arrays.asList(
// new Kruskal.Edge(0, 1, -3),
// new Kruskal.Edge(0, 4, 1),
// new Kruskal.Edge(4, 1, 4),
// new Kruskal.Edge(2, 1, 9),
// new Kruskal.Edge(2, 4, 3),
// new Kruskal.Edge(4, 3, 2)
// );
\end{verbatim}

\paragraph{Merge Sort}
\begin{verbatim}
// Implement a java test class called MergeSortTest using JUnit version 4.
// The class MergeSort that is to be tested contains the following:
// A public void mergeSort(int[]) method.
// Implement at least 3 different test cases.
\end{verbatim}

\paragraph{Quick Sort}
\begin{verbatim}
// Implement a java test class called QuickSortTest using JUnit version 4.
// The class QuickSort that is to be tested contains the following:
// A public void sort(int[]) method.
// Implement at least 3 different test cases.
\end{verbatim}

\subsubsection{Java - Book Example/AI Generated Reference Code}
\paragraph{Bellman Ford}
\begin{verbatim}
// Implement a java test class for the bellman ford algorithm above using JUnit version 4.
// Implement at least 3 different test cases.
// Build three graphs with the following edges and weights:
// Graph 1:
// List<BellmanFord.Edge> edges = Arrays.asList(
// new BellmanFord.Edge(0, 0, 1),
// new BellmanFord.Edge(1, 2, 3),
// new BellmanFord.Edge(1, 3, 2),
// new BellmanFord.Edge(1, 4, 2),
// new BellmanFord.Edge(3, 2, 5),
// new BellmanFord.Edge(4, 3, 3)
// );
// Graph 2:
// List<BellmanFord.Edge> edges = Arrays.asList(
// new BellmanFord.Edge(0, 1, -1),
// new BellmanFord.Edge(0, 2, 4),
// new BellmanFord.Edge(1, 2, 3),
// new BellmanFord.Edge(1, 4, -2),
// new BellmanFord.Edge(3, 2, 5),
// new BellmanFord.Edge(3, 1, 1),
// new BellmanFord.Edge(4, 3, -3)
// );
// Graph 3:
// List<BellmanFord.Edge> edges = Arrays.asList(
// new BellmanFord.Edge(0, 1, -1),
// new BellmanFord.Edge(0, 2, 4),
// new BellmanFord.Edge(1, 2, 3),
// new BellmanFord.Edge(1, 3, 2),
// new BellmanFord.Edge(1, 4, 2),
// new BellmanFord.Edge(3, 2, 5),
// new BellmanFord.Edge(3, 1, 1),
// new BellmanFord.Edge(4, 3, -3)
// );
\end{verbatim}

\paragraph{Binary Search}
\begin{verbatim}
// Implement a java test class for the binary search algorithm above using JUnit version 4.
// Implement at least 3 different test cases.
\end{verbatim}

\paragraph{Binary To Decimal}
\begin{verbatim}
// Implement a java test class for the binary to decimal algorithm above using JUnit version 4.
// Implement at least 3 different test cases.
\end{verbatim}

\paragraph{Breadth First Search}
\begin{verbatim}
// Implement a java test class for the breadth first search algorithm above using JUnit version 4.
// Implement at least 3 different test cases.
// Build a graph with the following edges:
// breadthFirstSearch.addEdge(0, 1);
// breadthFirstSearch.addEdge(0, 2);
// breadthFirstSearch.addEdge(1, 3);
// breadthFirstSearch.addEdge(2, 3);
// breadthFirstSearch.addEdge(3, 4);
// breadthFirstSearch.addEdge(3, 5);
\end{verbatim}

\paragraph{Depth First Search}
\begin{verbatim}
// Implement a java test class for the depth first search algorithm above using JUnit version 4.
// Implement at least 3 different test cases.
// Build a graph with the following edges:
// List<DepthFirstSearch.Edge> edges = Arrays.asList(
// new DepthFirstSearch.Edge(0, 1),
// new DepthFirstSearch.Edge(0, 3),
// new DepthFirstSearch.Edge(1, 4),
// new DepthFirstSearch.Edge(2, 4),
// new DepthFirstSearch.Edge(2, 5),
// new DepthFirstSearch.Edge(3, 1),
// new DepthFirstSearch.Edge(4, 3),
// new DepthFirstSearch.Edge(5, 5),
// new DepthFirstSearch.Edge(6, 6)
// );
\end{verbatim}

\paragraph{Dijkstra}
\begin{verbatim}
// Implement a java test class for the dijkstra algorithm above using JUnit version 4.
// Implement at least 3 different test cases.
// Build three graphs with the following edges and weights:
// Graph 1:
// List<Dijkstra.Edge> edges = Arrays.asList(
// new Dijkstra.Edge(0, 1, 1),
// new Dijkstra.Edge(0, 2, 4),
// new Dijkstra.Edge(1, 2, 3),
// new Dijkstra.Edge(1, 3, 2),
// new Dijkstra.Edge(1, 4, 2),
// new Dijkstra.Edge(3, 2, 5),
// new Dijkstra.Edge(4, 3, 3)
// );
// Graph 2:
// List<Dijkstra.Edge> edges = Arrays.asList(
// new Dijkstra.Edge(0, 1, 1),
// new Dijkstra.Edge(0, 2, 4),
// new Dijkstra.Edge(1, 2, 3),
// new Dijkstra.Edge(1, 3, 2),
// new Dijkstra.Edge(1, 4, 2),
// new Dijkstra.Edge(3, 2, 5),
// new Dijkstra.Edge(3, 1, 1),
// new Dijkstra.Edge(4, 3, 3)
// );
// Graph 3:
// List<Dijkstra.Edge> edges = Arrays.asList(
// new Dijkstra.Edge(0, 0, 1),
// new Dijkstra.Edge(1, 2, 3),
// new Dijkstra.Edge(1, 3, 2),
// new Dijkstra.Edge(1, 4, 2),
// new Dijkstra.Edge(3, 2, 5),
// new Dijkstra.Edge(4, 3, 3)
// );
\end{verbatim}

\paragraph{Egyptian Fractions}
\begin{verbatim}
// Implement a java test class for the egyptian fractions algorithm above using JUnit version 4.
// Implement at least 3 different test cases.
\end{verbatim}

\paragraph{Floyd Warshall}
\begin{verbatim}
// Implement a java test class for the floyd warshall algorithm above using JUnit version 4.
// Implement at least 3 different test cases.
// Build two graphs with the following edges and weights:
// Graph 1:
// floydWarshall.addEdge(0, 1, 10);
// floydWarshall.addEdge(0, 3, 5);
// floydWarshall.addEdge(1, 3, 2);
// floydWarshall.addEdge(1, 2, 1);
// floydWarshall.addEdge(2, 4, 4);
// floydWarshall.addEdge(3, 1, 3);
// floydWarshall.addEdge(3, 2, 9);
// floydWarshall.addEdge(3, 4, 2);
// floydWarshall.addEdge(4, 2, 6);
// Graph 2:
// floydWarshall.addEdge(0, 2, 5);
// floydWarshall.addEdge(1, 3, 5);
// floydWarshall.addEdge(2, 3, 4);
// floydWarshall.addEdge(3, 2, 4);
// floydWarshall.addEdge(4, 3, 4);
// floydWarshall.addEdge(2, 4, -2);
\end{verbatim}

\paragraph{Knapsack}
\begin{verbatim}
// Implement a java test class for the knapsack algorithm above using JUnit version 4.
// Implement at least 3 different test cases.
\end{verbatim}

\paragraph{Kruskal}
\begin{verbatim}
// Implement a java test class for the kruskal algorithm above using JUnit version 4.
// Implement at least 3 different test cases.
// Build two graphs with the following edges and weights:
// Graph 1:
// List<Kruskal.Edge> edges = Arrays.asList(
// new Kruskal.Edge(0, 1, 7),
// new Kruskal.Edge(1, 2, 8),
// new Kruskal.Edge(0, 3, 5),
// new Kruskal.Edge(1, 3, 9),
// new Kruskal.Edge(1, 4, 7),
// new Kruskal.Edge(2, 4, 5),
// new Kruskal.Edge(3, 4, 15),
// new Kruskal.Edge(3, 5, 6),
// new Kruskal.Edge(4, 5, 8),
// new Kruskal.Edge(4, 6, 9),
// new Kruskal.Edge(5, 6, 11)
// );
// Graph 2:
// List<Kruskal.Edge> edges = Arrays.asList(
// new Kruskal.Edge(0, 1, -3),
// new Kruskal.Edge(0, 4, 1),
// new Kruskal.Edge(4, 1, 4),
// new Kruskal.Edge(2, 1, 9),
// new Kruskal.Edge(2, 4, 3),
// new Kruskal.Edge(4, 3, 2)
// );
\end{verbatim}

\paragraph{Merge Sort}
\begin{verbatim}
// Implement a java test class for the merge sort algorithm above using JUnit version 4.
// Implement at least 3 different test cases.
\end{verbatim}

\paragraph{Quick Sort}
\begin{verbatim}
// Implement a java test class for the quick sort algorithm above using JUnit version 4.
// Implement at least 3 different test cases.
\end{verbatim}
\subsubsection{Python - Prompt Only}
\paragraph{Bellman Ford}
\begin{verbatim}
# Implement a python test class called BellmanFordTest using the unittest module.
# The class BellmanFord that is to be tested contains the following methods:
# A constructor, which takes an integer "vertices".
# A public method add_edge(), which takes three integers "source", "destination", "weight" and returns nothing.
# A public method bellman_ford(), which takes an integer "source" and returns a dictionary, 
containing the shortest path to every other node in the graph.
# If there is a negative cycle, the method returns None.
# If there is no path from source to destination or if the source and destination node are 
the same, this node is skipped in the output.
# Implement at least 3 different test cases.
# Build three graphs with the following edges and weights:
# Graph 1:
# self.bellmanFord.add_edge(0, 1, -1)
# self.bellmanFord.add_edge(0, 2, 4)
# self.bellmanFord.add_edge(1, 2, 3)
# self.bellmanFord.add_edge(1, 3, 2)
# self.bellmanFord.add_edge(1, 4, 2)
# self.bellmanFord.add_edge(3, 1, 1)
# self.bellmanFord.add_edge(4, 3, -3)
# Graph 2:
# self.bellmanFord.add_edge(0, 1, -1)
# self.bellmanFord.add_edge(0, 2, 4)
# self.bellmanFord.add_edge(1, 2, 3)
# self.bellmanFord.add_edge(1, 4, -2)
# self.bellmanFord.add_edge(3, 2, 5)
# self.bellmanFord.add_edge(3, 1, 1)
# self.bellmanFord.add_edge(4, 3, -3)
# Graph 3:
# self.bellmanFord.add_edge(0, 0, 1)
# self.bellmanFord.add_edge(1, 2, 3)
# self.bellmanFord.add_edge(1, 3, 2)
# self.bellmanFord.add_edge(1, 4, 2)
# self.bellmanFord.add_edge(3, 2, 5)
# self.bellmanFord.add_edge(4, 3, 3)
\end{verbatim}

\paragraph{Binary Search}
\begin{verbatim}
# Implement a python test class called BinarySearchTest using the unittest module.
# The class BinarySearch that is to be tested contains the following methods:
# A public method binary_search(), which takes an integer and an integer list as parameters 
and returns a boolean.
# Implement at least 3 different test cases.
\end{verbatim}

\paragraph{Binary To Decimal}
\begin{verbatim}
# Implement a python test class called BinaryToDecimalTest using the unittest module.
# The class BinaryToDecimal that is to be tested contains the following methods:
# A public method convert_to_decimal(), which takes a string "binary" as parameter and 
returns an integer with the decimal.
# Implement at least 3 different test cases.
\end{verbatim}

\paragraph{Breadth First Search}
\begin{verbatim}
# Implement a python test class called GraphTest using the unittest module.
# The class Graph that is to be tested implements the breadth first search algorithm and 
contains the following methods:
# A constructor that takes an integer "vertices".
# A public method bfs(), which takes an integer "vertex" as parameter and returns an integer 
list of visited nodes in the graph.
# A public method add_edge(), which takes two integers as parameters and adds an edge between two vertices.
# Implement at least 3 different test cases.
# Build a graph with the following edges:
# self.graph.add_edge(0, 1)
# self.graph.add_edge(1, 0)
# self.graph.add_edge(0, 2)
# self.graph.add_edge(2, 0)
# self.graph.add_edge(1, 3)
# self.graph.add_edge(3, 1)
# self.graph.add_edge(2, 3)
# self.graph.add_edge(3, 2)
# self.graph.add_edge(3, 4)
# self.graph.add_edge(4, 3)
# self.graph.add_edge(3, 5)
# self.graph.add_edge(5, 3)
\end{verbatim}

\paragraph{Depth First Search}
\begin{verbatim}
# Implement a python test class called DepthFirstSearchTest using the unittest module.
# The class DepthFirstSearch that is to be tested contains the following methods:
# A public method add_edge(), which takes two integers "source" and "destination" as 
parameters and returns nothing.
# A public method dfs(), which takes an integer "start_node" and returns a list of 
visited nodes.
# Consider all possible permutations.
# Implement at least 3 different test cases.
# Build a graph with the following edges:
# self.depthFirstSearch.add_edge(0, 1)
# self.depthFirstSearch.add_edge(1, 0)
# self.depthFirstSearch.add_edge(0, 3)
# self.depthFirstSearch.add_edge(3, 0)
# self.depthFirstSearch.add_edge(1, 4)
# self.depthFirstSearch.add_edge(4, 1)
# self.depthFirstSearch.add_edge(2, 4)
# self.depthFirstSearch.add_edge(4, 2)
# self.depthFirstSearch.add_edge(2, 5)
# self.depthFirstSearch.add_edge(5, 2)
# self.depthFirstSearch.add_edge(3, 1)
# self.depthFirstSearch.add_edge(1, 3)
# self.depthFirstSearch.add_edge(4, 3)
# self.depthFirstSearch.add_edge(3, 4)
# self.depthFirstSearch.add_edge(5, 5)
# self.depthFirstSearch.add_edge(6, 6)
\end{verbatim}

\paragraph{Dijkstra}
\begin{verbatim}
# Implement a python test class called DijkstraTest using the unittest module.
# The class Dijkstra that is to be tested contains the following methods:
# A constructor, which takes an integer "vertices".
# A public method add_edge(), which takes three integers "source", "destination", "weight" and returns nothing.
# A public method find_shortest_paths(), which takes an integer "source" and returns a
dictionary, containing the shortest path to every other node in the graph.
# If there is no path from source to destination or if the source and destination node 
are the same, this node is skipped in the output.
# Implement at least 3 different test cases.
# Build two graphs with the following edges and weights:
# Graph 1:
# self.dijkstra.add_edge(0, 1, 1)
# self.dijkstra.add_edge(0, 2, 4)
# self.dijkstra.add_edge(1, 2, 3)
# self.dijkstra.add_edge(1, 3, 2)
# self.dijkstra.add_edge(1, 4, 2)
# self.dijkstra.add_edge(3, 2, 5)
# self.dijkstra.add_edge(3, 1, 1)
# self.dijkstra.add_edge(4, 3, 3)
# Graph 2:
# self.dijkstra.add_edge(0, 0, 1)
# self.dijkstra.add_edge(1, 2, 3)
# self.dijkstra.add_edge(1, 3, 2)
# self.dijkstra.add_edge(1, 4, 2)
# self.dijkstra.add_edge(3, 2, 5)
# self.dijkstra.add_edge(4, 3, 3)
\end{verbatim}

\paragraph{Egyptian Fractions}
\begin{verbatim}
# Implement a python test class called EgyptianFractionsTest using the unittest module.
# The class EgyptianFractions that is to be tested contains the following methods:
# A public method build(), which takes two long "numerator" and "denominator" as parameters and returns a list.
# Implement at least 3 different test cases.
\end{verbatim}

\paragraph{Floyd Warshall}
\begin{verbatim}
# Implement a python test class called FloydWarshallTest using the unittest module.
# The class FloydWarshall that is to be tested contains the following methods:
# A constructor, which takes an integer "nodes".
# A public method add_edge(), which takes three integers "source", "destination", "weight"
as parameters and returns nothing.
# A public method path(), which takes two integers "source" and "destination" and returns
an integer list.
# A public method run(), which runs the floyd warshall algorithm.
# If there is no path from source to destination or if the source and destination node are 
the same, this node is skipped in the output.
# Implement at least 3 different test cases.
# Build two graphs with the following edges and weights:
# Graph 1:
# self.floydWarshall.add_edge(0, 1, 10)
# self.floydWarshall.add_edge(0, 3, 5)
# self.floydWarshall.add_edge(1, 3, 2)
# self.floydWarshall.add_edge(1, 2, 1)
# self.floydWarshall.add_edge(2, 4, 4)
# self.floydWarshall.add_edge(3, 1, 3)
# self.floydWarshall.add_edge(3, 2, 9)
# self.floydWarshall.add_edge(3, 4, 2)
# self.floydWarshall.add_edge(4, 2, 6)
# Graph 2:
# self.floydWarshall.add_edge(0, 2, 5)
# self.floydWarshall.add_edge(1, 3, 5)
# self.floydWarshall.add_edge(2, 3, 4)
# self.floydWarshall.add_edge(3, 2, 4)
# self.floydWarshall.add_edge(4, 3, 4)
# self.floydWarshall.add_edge(2, 4, -2)
\end{verbatim}

\paragraph{Knapsack}
\begin{verbatim}
# Implement a python test class called KnapsackTest using the unittest module.
# The class Knapsack that is to be tested contains the following methods:
# A public method bottom_up(), which takes an integer "capacity", an integer list "weights"
and an integer list "values" as parameters and returns an integer with the maximum value.
# Implement at least 3 different test cases.
\end{verbatim}

\paragraph{Kruskal}
\begin{verbatim}
# Implement a python test class called KruskalTest using the unittest module.
# The class Kruskal that is to be tested contains the following methods:
# A constructor, which takes an integer "vertices".
# A public method add_edge(), which takes three integers "source", "destination", "weight"
as parameters and returns nothing.
# A public method run_kruskal_algorithm(), which creates the minimal spanning tree and returns a list of edges.
# Implement at least 3 different test cases.
# Build two graphs with the following edges and weights:
# Graph 1:
# self.kruskal.add_edge(5, 6, 11)
# self.kruskal.add_edge(0, 1, 7)
# self.kruskal.add_edge(1, 2, 8)
# self.kruskal.add_edge(0, 3, 5)
# self.kruskal.add_edge(1, 3, 9)
# self.kruskal.add_edge(1, 4, 7)
# self.kruskal.add_edge(2, 4, 5)
# self.kruskal.add_edge(3, 4, 15)
# self.kruskal.add_edge(3, 5, 6)
# self.kruskal.add_edge(4, 5, 8)
# self.kruskal.add_edge(4, 6, 9)
# Graph 2:
# self.kruskal.add_edge(0, 1, -3)
# self.kruskal.add_edge(0, 4, 1)
# self.kruskal.add_edge(4, 1, 4)
# self.kruskal.add_edge(2, 1, 9)
# self.kruskal.add_edge(2, 4, 3)
# self.kruskal.add_edge(4, 3, 2)
\end{verbatim}

\paragraph{Merge Sort}
\begin{verbatim}
# Implement a python test class called MergeSortTest using the unittest module.
# The class MergeSort that is to be tested contains the following methods:
# A public method merge_sort(), which takes an integer list as parameter and returns a sorted integer list.
# Implement at least 3 different test cases.
\end{verbatim}

\paragraph{Quick Sort}
\begin{verbatim}
# Implement a python test class called QuickSortTest using the unittest module.
# The class QuickSort that is to be tested contains the following methods:
# A public method sort(), which takes an integer list as parameter and returns nothing.
# Implement at least 3 different test cases.
\end{verbatim}

\subsubsection{Python - Book Example/AI Generated Reference Code}
\paragraph{Bellman Ford}
\begin{verbatim}
# Implement a test class for the bellman ford algorithm above using the python unittest module.
# Implement at least 3 different test cases.
# Build three graphs with the following edges and weights:
# Graph 1:
# self.bellmanFord.add_edge(0, 1, -1)
# self.bellmanFord.add_edge(0, 2, 4)
# self.bellmanFord.add_edge(1, 2, 3)
# self.bellmanFord.add_edge(1, 3, 2)
# self.bellmanFord.add_edge(1, 4, 2)
# self.bellmanFord.add_edge(3, 1, 1)
# self.bellmanFord.add_edge(4, 3, -3)
# Graph 2:
# self.bellmanFord.add_edge(0, 1, -1)
# self.bellmanFord.add_edge(0, 2, 4)
# self.bellmanFord.add_edge(1, 2, 3)
# self.bellmanFord.add_edge(1, 4, -2)
# self.bellmanFord.add_edge(3, 2, 5)
# self.bellmanFord.add_edge(3, 1, 1)
# self.bellmanFord.add_edge(4, 3, -3)
# Graph 3:
# self.bellmanFord.add_edge(0, 0, 1)
# self.bellmanFord.add_edge(1, 2, 3)
# self.bellmanFord.add_edge(1, 3, 2)
# self.bellmanFord.add_edge(1, 4, 2)
# self.bellmanFord.add_edge(3, 2, 5)
# self.bellmanFord.add_edge(4, 3, 3)
\end{verbatim}

\paragraph{Binary Search}
\begin{verbatim}
# Implement a test class for the binary search algorithm above using the python unittest module.
# Implement at least 3 different test cases.
\end{verbatim}

\paragraph{Binary To Decimal}
\begin{verbatim}
# Implement a test class for the binary to decimal algorithm above using the python unittest module.
# Implement at least 3 different test cases.
\end{verbatim}

\paragraph{Breadth First Search}
\begin{verbatim}
# Implement a test class for the breadth first search algorithm above using the python unittest module.
# Implement at least 3 different test cases.
# Build a graph with the following edges:
# self.graph.add_edge(0, 1)
# self.graph.add_edge(1, 0)
# self.graph.add_edge(0, 2)
# self.graph.add_edge(2, 0)
# self.graph.add_edge(1, 3)
# self.graph.add_edge(3, 1)
# self.graph.add_edge(2, 3)
# self.graph.add_edge(3, 2)
# self.graph.add_edge(3, 4)
# self.graph.add_edge(4, 3)
# self.graph.add_edge(3, 5)
# self.graph.add_edge(5, 3)
\end{verbatim}

\paragraph{Depth First Search}
\begin{verbatim}
# Implement a test class for the depth first search algorithm above using the python unittest module.
# Implement at least 3 different test cases.
# Consider all possible permutations.
# Build a graph with the following edges:
# self.depthFirstSearch.add_edge(0, 1)
# self.depthFirstSearch.add_edge(1, 0)
# self.depthFirstSearch.add_edge(0, 3)
# self.depthFirstSearch.add_edge(3, 0)
# self.depthFirstSearch.add_edge(1, 4)
# self.depthFirstSearch.add_edge(4, 1)
# self.depthFirstSearch.add_edge(2, 4)
# self.depthFirstSearch.add_edge(4, 2)
# self.depthFirstSearch.add_edge(2, 5)
# self.depthFirstSearch.add_edge(5, 2)
# self.depthFirstSearch.add_edge(3, 1)
# self.depthFirstSearch.add_edge(1, 3)
# self.depthFirstSearch.add_edge(4, 3)
# self.depthFirstSearch.add_edge(3, 4)
# self.depthFirstSearch.add_edge(5, 5)
# self.depthFirstSearch.add_edge(6, 6)
\end{verbatim}

\paragraph{Dijkstra}
\begin{verbatim}
# Implement a test class for the dijkstra algorithm above using the python unittest module.
# Implement at least 3 different test cases.
# Build two graphs with the following edges and weights:
# Graph 1:
# self.dijkstra.add_edge(0, 1, 1)
# self.dijkstra.add_edge(0, 2, 4)
# self.dijkstra.add_edge(1, 2, 3)
# self.dijkstra.add_edge(1, 3, 2)
# self.dijkstra.add_edge(1, 4, 2)
# self.dijkstra.add_edge(3, 2, 5)
# self.dijkstra.add_edge(3, 1, 1)
# self.dijkstra.add_edge(4, 3, 3)
# Graph 2:
# self.dijkstra.add_edge(0, 0, 1)
# self.dijkstra.add_edge(1, 2, 3)
# self.dijkstra.add_edge(1, 3, 2)
# self.dijkstra.add_edge(1, 4, 2)
# self.dijkstra.add_edge(3, 2, 5)
# self.dijkstra.add_edge(4, 3, 3)
\end{verbatim}

\paragraph{Egyptian Fractions}
\begin{verbatim}
# Implement a test class for the egyptian fractions algorithm above using the python unittest module.
# Implement at least 3 different test cases.
\end{verbatim}

\paragraph{Floyd Warshall}
\begin{verbatim}
# Implement a test class for the floyd warshall algorithm above using the python unittest module.
# Implement at least 3 different test cases.
# Build two graphs with the following edges and weights:
# Graph 1:
# self.floydWarshall.add_edge(0, 1, 10)
# self.floydWarshall.add_edge(0, 3, 5)
# self.floydWarshall.add_edge(1, 3, 2)
# self.floydWarshall.add_edge(1, 2, 1)
# self.floydWarshall.add_edge(2, 4, 4)
# self.floydWarshall.add_edge(3, 1, 3)
# self.floydWarshall.add_edge(3, 2, 9)
# self.floydWarshall.add_edge(3, 4, 2)
# self.floydWarshall.add_edge(4, 2, 6)
# Graph 2:
# self.floydWarshall.add_edge(0, 2, 5)
# self.floydWarshall.add_edge(1, 3, 5)
# self.floydWarshall.add_edge(2, 3, 4)
# self.floydWarshall.add_edge(3, 2, 4)
# self.floydWarshall.add_edge(4, 3, 4)
# self.floydWarshall.add_edge(2, 4, -2)
\end{verbatim}

\paragraph{Knapsack}
\begin{verbatim}
# Implement a test class for the knapsack algorithm above using the python unittest module.
# Implement at least 3 different test cases.
\end{verbatim}

\paragraph{Kruskal}
\begin{verbatim}
# Implement a test class for the kruskal algorithm above using the python unittest module.
# Implement at least 3 different test cases.
# Build two graphs with the following edges and weights:
# Graph 1:
# self.kruskal.add_edge(5, 6, 11)
# self.kruskal.add_edge(0, 1, 7)
# self.kruskal.add_edge(1, 2, 8)
# self.kruskal.add_edge(0, 3, 5)
# self.kruskal.add_edge(1, 3, 9)
# self.kruskal.add_edge(1, 4, 7)
# self.kruskal.add_edge(2, 4, 5)
# self.kruskal.add_edge(3, 4, 15)
# self.kruskal.add_edge(3, 5, 6)
# self.kruskal.add_edge(4, 5, 8)
# self.kruskal.add_edge(4, 6, 9)
# Graph 2:
# self.kruskal.add_edge(0, 1, -3)
# self.kruskal.add_edge(0, 4, 1)
# self.kruskal.add_edge(4, 1, 4)
# self.kruskal.add_edge(2, 1, 9)
# self.kruskal.add_edge(2, 4, 3)
# self.kruskal.add_edge(4, 3, 2)
\end{verbatim}

\paragraph{Merge Sort}
\begin{verbatim}
# Implement a test class for the merge sort algorithm above using the python unittest module.
# Implement at least 3 different test cases.
\end{verbatim}

\paragraph{Quick Sort}
\begin{verbatim}
# Implement a test class for the quicksort algorithm above using the python unittest module.
# Implement at least 3 different test cases.
\end{verbatim}

\end{document}